\documentclass[journal,twoside,web]{ieeecolor}
\usepackage{generic}
\usepackage{cite}
\usepackage{amsmath,amssymb,amsfonts}
\usepackage{algorithmic}
\usepackage{graphicx}
\usepackage{textcomp}

\usepackage{units} 
\usepackage{multirow}
\usepackage{rotating}
\usepackage{lettrine}

\usepackage{hyperref}

\usepackage{bm}
\usepackage[acronym]{glossaries}

\usepackage[table]{xcolor}

\newcommand{\Bg}{\cellcolor[HTML]{C0F8C0}}
\newcommand{\Br}{\cellcolor[HTML]{E0C0C0}}

\usepackage{xcolor}

\usepackage{alltt}
\usepackage{tikz}
\usetikzlibrary{shapes, arrows.meta, positioning, calc}

\newcommand*\toeast[1]{\tikz[baseline=(char.base)]{\node[shape=signal, signal to=east,draw,inner sep=1pt] (char) {#1};}}

\usepackage[pscoord]{eso-pic}
\newcommand{\placetextbox}[3]{%
  \setbox0=\hbox{#3}
  \AddToShipoutPictureFG*{
    \put(\LenToUnit{#1\paperwidth},\LenToUnit{#2\paperheight}){\vtop{{\null}\makebox[0pt][c]{#3}}}%
  }%
}%

\newacronym{fdr}{FDR}{frame delivery ratio}

\def\BibTeX{{\rm B\kern-.05em{\sc i\kern-.025em b}\kern-.08em
    T\kern-.1667em\lower.7ex\hbox{E}\kern-.125emX}}
\author{\makebox[\textwidth][c]{Gabriele Formis, Gianluca Cena, Lukasz Wisniewski, Stefano Scanzio}}

\begin{document}
\placetextbox{0.5}{1}{This is the author's version of an article accepted to IEEE Transactions on Industrial Informatics.}
\placetextbox{0.5}{0.985}{Changes were made to this version by the publisher prior to publication.}
\placetextbox{0.5}{0.97}{The final version of record is available at \href{https://doi.org/10.1109/TII.2025.3609224}{https://doi.org/10.1109/TII.2025.3609224}}%
\placetextbox{0.5}{0.05}{Copyright (c) 2025 IEEE. Personal use is permitted.}
\placetextbox{0.5}{0.035}{For any other purposes, permission must be obtained from the IEEE by emailing pubs-permissions@ieee.org.}%

\title{Accurate and Efficient Prediction of \mbox{Wi-Fi} Link Quality Based on Machine Learning
\thanks{This work was partially supported by the European Union under the Italian National Recovery and Resilience Plan (NRRP) of NextGenerationEU, partnership on ``Telecommunications of the Future'' (PE00000001 - program ``RESTART''). (Corresponding author: Stefano Scanzio.)\\
Gabriele Formis is with Politecnico di Torino, Italy, and CNR-IEIIT, National Research Council of Italy, Italy (e-mail: gabriele.formis@polito.it.) Gianluca Cena and Stefano Scanzio are with CNR-IEIIT, National Research Council of Italy, Italy (e-mail: gianluca.cena@cnr.it, stefano.scanzio@cnr.it.) Lukasz Wisniewski is with the Institute Industrial IT - inIT, Technische Hochschule OWL, Germany (e-mail: lukasz.wisniewski@th-owl.de.)
}}

\maketitle

\begin{abstract}

Wireless communications are characterized by their unpredictability, posing challenges for maintaining consistent communication quality. 
This paper presents a comprehensive analysis of various prediction models, with a focus on achieving accurate and efficient Wi-Fi link quality forecasts using machine learning techniques. 
Specifically, the paper evaluates the performance of data-driven models based on the linear combination of exponential moving averages, which are designed for low-complexity implementations and are then suitable for hardware platforms with limited processing resources.

Accuracy of the proposed approaches was assessed using experimental data from a real-world Wi-Fi testbed, considering both channel-dependent and channel-independent training data.
Remarkably, channel-independent models, which allow for generalized training by equipment manufacturers, demonstrated competitive performance.
Overall, this
study provides 
insights into the practical deployment of machine learning-based prediction models for enhancing Wi-Fi dependability in industrial environments.
\end{abstract}

\begin{IEEEkeywords}
Data-driven Models, Wi-Fi, Channel Quality Prediction, EMA, Machine Learning
\end{IEEEkeywords}


\section{Introduction}
\label{sec:introduction}
\lettrine[lines=2, findent=0.2em, nindent=0em]{W}ireless
communications suffer from disturbance due to the open nature of the transmission medium, including interference from nearby communication equipment operating in the same frequency range and electromagnetic noise generated by power equipment, which may abound in industrial plants \cite{7299315}.
These phenomena impact on frame transmission attempts tangibly, to the point that their outcomes (either success or failure) can be modeled as binary random variables, whose probability varies over time since spectrum conditions are typically non-stationary.
This leads to unreliable behavior, which cannot be tolerated by applications.
For this reason, Automatic Repeat-reQuest (ARQ) mechanisms are customarily included in the media access control (MAC) layer, which allow to cope with frame losses by exploiting confirmations and retransmissions, even though doing so enlarges communication latency and jitters.

As a matter of fact, the lack of adequate determinism affects most of the existing wireless network technologies, slowing down their adoption in industrial environments \cite{10070395}.
Consequently, communication for automation at present still relies for the most part on cables, and is progressively migrating from legacy solutions (fieldbuses) to those based on IEEE 802.3 (industrial Ethernet), which according to the yearly HMS Study 
has grown by $12\%$ 
and now reaches $71\%$ 
of the installed base \cite{hms24}.
The same study also shows that wireless technologies have seen a steady growth in recent years and constitute $7\%$ of the total share in 2024.

The increase in the usage of wireless network technologies in industrial environments 
is partially motivated by practicability, particularly in the Industry 4.0 context \cite{KUNST20191}, whenever
wiring is either impossible or simply inconvenient.
This is the case, e.g., of automated guided vehicle (AGV) and autonomous mobile robots (AMR) \cite{fragapaneIncreasingFlexibilityProductivity2022}, which must coordinate themselves in real-time and integrate their operations in the overall production process.
The two most appealing
solutions to 
provide wireless extensions of wired 
infrastructures in 
such scenarios \cite{10372393}  
are probably IEEE 802.11  
\cite{IEEE80211_2024},
also known as \mbox{Wi-Fi}, and mobile 5G networks.
Other solutions exist for industry, e.g., IO-Link Wireless (based on the IEEE 802.15.1 PHY) and WirelessHART (based on IEEE 802.15.4), but they offer a noticeably lower throughput and are unsuitable for applications like industrial augmented reality \cite{2020-IEEE-ACC-WiFi5G}.
Wi-Fi is the only technology that
ensures direct interoperability 
with Ethernet at the data-link layer,
as both 
rely on the same addressing space (EUI-48) and have similar maximum transmission units (MTU)
and speed.
Moreover, the acquisition and maintenance costs for Wi-Fi are noticeably lower than  
private 5G networks, which are also likely to include licensing fees.
Hence, \mbox{Wi-Fi} is preferable for indoor use (shop-floor, warehouse), 
as witnessed by commercial solutions based on modified \mbox{Wi-Fi~4} technology like Siemens' iPCF and WIA-FA \cite{62948},
whereas 5G ensures ubiquitous connectivity.

Making \mbox{Wi-Fi} behavior more dependable and deterministic has been one of the main research goals of the past two decades.
The distributed coordination function (DCF) and the enhanced distributed channel access (EDCA) of the hybrid coordination function (HCF) are essentially random mechanisms.
Centralized solutions, like the point coordination function (PCF), the HCF controlled channel access (HCCA), and more recently the much more promising trigger frames in \mbox{Wi-Fi}~6, although able to enhance determinism, are unsuitable to tackle sources of disturbance other than nearby \mbox{Wi-Fi} devices.

Improving Wi-Fi dependability constitutes the primary focus of the IEEE 802.11bn Ultra High Reliability (UHR) Task Group, which is in charge to define \mbox{Wi-Fi} 8 \cite{10634004}.
Among the many mechanisms that can be exploited for this purpose, link-level seamless redundancy (Wi-Red) \cite{TII2016} and coordinated spatial reuse \cite{2023-CSCN-CSR} are deemed promising options.
In addition, machine learning (ML) techniques can be also exploited to deliver better features.
In particular, the ability to foresee with acceptable confidence the behavior of the wireless spectrum in the near future may enable mechanisms aimed at improving link quality \cite{9945847}.
Such mechanisms can operate at the MAC layer, in which case the timeframe for forecasts is short (milliseconds up to a second), and at the application layer (including network management), where prediction intervals are larger (seconds to minutes).
For example, both rate adaptation (e.g., Minstrel) and packet steering (in Wi-Fi~7) algorithms can benefit
from accurate predictions.

Practical feasibility and cost are essential aspects that impact on the chance of any proposal to be considered for inclusion in commercial devices.
For this reason, in this paper we seek for prediction models that feature adequate accuracy subject to the constraint that they must allow for efficient implementation on hardware platforms with limited processing resources (CPU and memory), e.g., access points (AP) and wireless adapters for end stations (STA).
In particular, we focused on methods that rely on the exponentially weighted moving average (EMA), which is known to enable fast and lightweight implementations, incorporating modifications aimed at maximizing the ability to accurately predict link quality by minimizing the mean prediction error.

The models we present here are data-driven, that is, their operation depends on parameters whose value is determined through a suitable training phase carried out on data acquired from the real world.
As we show, doing so yields 
tangible improvements in terms of prediction accuracy.
We also note that results are not completely specific to the environment in which the link quality measurements are performed.
Conversely, they are of broader validity, meaning that training was able to capture more general characteristics about the disturbance affecting real \mbox{Wi-Fi} communications,
not directly related to the considered scenario.

The paper is structured as follows:
in Section~\ref{sec:SOTA} a brief survey is provided about the state of the art on the subject, 
while in Section~\ref{sec:PRED} three simple models are presented for predicting link quality.
Results of a post-analysis carried out on experimental data are reported in Section~\ref{sec:RESULT},
whereas some conclusions are drawn in Section~\ref{sec:CONC}.

\section{State of the Art}
\label{sec:SOTA}

Industrial wireless communication systems are continually evolving, driven by the ever increasing demand for high-speed, low-latency, deterministic and reliable connectivity in several application fields, including AGVs, mobile 
human-machine interface (HMI) 
devices, and Internet of Things (IoT) sensors, to cite a few. 
This evolution has led to a pressing need for more sophisticated techniques to make communication protocols able to adapt to the dynamic nature of the wireless spectrum \cite{9452149, 8884240, 8813020}.
One notable challenge is the unpredictability of wireless channel quality, which may fluctuate due to a multitude of factors, including environmental conditions, interference from other devices, and mobility of users and objects, which pose a significant obstacle in achieving consistent and efficient data transmission \cite{10213404}.

While traditional methods for predicting channel quality have served as the foundation for early wireless networks, they often fall short in capturing the intricacies of real-world scenarios. 
This is due to the fact that they typically rely on simplified statistical models, which may not adequately account for the actual nature of the wireless spectrum. 
As the spectrum becomes increasingly crowded, with numerous nodes in motion, the wireless environment becomes particularly challenging when it comes to coexistence issues.  
In contrast, the emergence of ML has revolutionized the approach to wireless channel prediction (WCP) \cite{10295470, 2022-ITL-ML, 9618666, 9924163}. 
ML models, particularly artificial neural networks (ANN) and deep neural networks, have demonstrated remarkable capabilities in learning complex patterns and relationships from large datasets 
\cite{8761934, 9665388, 9887796, 10422880}.
By analyzing historical channel data, these models can extract valuable insights and make accurate predictions about future channel conditions.

Recent research has explored the application of deep learning (DL), and ML in general, to a number of contexts specifically related to industrial networks, 
such as security \cite{10570153}, 
network resource allocation \cite{9976231}, but also WCP techniques for both \mbox{Wi-Fi} \cite{9786784} and 6G \cite{10129974,9795904}, 
with promising results. 
In \cite{9165822} a DL-based approach is proposed for channel prediction in millimeter-wave communication systems.
Similarly, in \cite{8395053} a convolutional neural network (CNN) model is developed for predicting channel quality in dynamic vehicular environments, showing the potential of DL in addressing the challenges of wireless communication.

The high heterogeneity of traffic, environment, nodes, and protocols is one of the reasons to exploit ML for WCP. 
ML can be profitably applied to Wi-Fi in a wide range of application contexts, to predict different relevant quantities and optimize a number of KPIs.
In \cite{KHAN2020102499} throughput is predicted by means of a data-driven approach based on several ML methods. 
Handover prediction and AP selection problems are the main goals addressed in \cite{Khan2022}.
Instead, \cite{10314514} focuses on the reduction of power consumption through reinforcement learning, a technique commonly used for optimization (e.g., rate adaptation in Wi-Fi) but hardly suitable for WCP. 
The problem of the coexistence among network protocols in the \unit[2.4]{GHz} band (e.g., Wi-Fi and Bluetooth) is tackled in \cite{9217227}, with the goal of optimizing the spectrum usage.

In this work the prediction target is the future frame delivery ratio (FDR), but
other link quality metrics exist as well, like the 
received signal strength indicator (RSSI)
or the channel occupancy.
We opted for the FDR as its forecasts can be exploited by the protocol layers above the radio block,
e.g., to try preventing latency-related constraints of industrial applications from being missed.

Unfortunately, the adoption of complex ML models comes with its own set of challenges, particularly in terms of computation time and resource requirements.
Deploying these models on resource-constrained devices, like IoT sensors and low-power wireless nodes, may prove to be impractical due to limitations in processing power and memory.
To address them, alternative approaches are being explored that strike a balance between prediction accuracy and computational effort. 
One promising direction are lightweight ML models that can run efficiently on embedded hardware platforms \cite{9149180}. 
These models leverage techniques such as quantization, pruning, and model compression to reduce their memory and computational footprint, while maintaining satisfactory performance.

In summary, the search for reliable and efficient wireless communication systems continues to drive innovation in WCP. 
While advanced ML techniques have unlocked new possibilities for achieving higher levels of accuracy and adaptability, their practical implementation requires careful consideration of both computational constraints and deployment efforts in real-world scenarios.

\section{Prediction models for \mbox{Wi-Fi} link quality}
\label{sec:PRED}
This work considers a wireless link between a pair of nodes, e.g., the AP and an associated STA in a \mbox{Wi-Fi} infrastructure network.
However, the analysis can be easily applied to other wireless communication technologies and scenarios as well.
To characterize link quality, a cyclic probing is carried out by means of confirmed one-shot frame transmissions (no retries allowed).
The receiver (the AP, in this case) confirms the correct reception of the $i$-th data frame by returning an acknowledgment (ACK) frame. 
Depending on whether the ACK frame correctly came back to the sender or not, 
the outcome $x_i=1$ (\textit{success}) or $x_i=0$ (\textit{failure}) 
were logged, respectively. 
For every experiment a dataset is created that consists of the ordered sequence of outcomes 
$\mathcal{D} = \left( x_1, \cdots, x_i, \cdots, x_{|\mathcal{D}|} \right)$
obtained from the testbed on a time interval long enough to provide significant statistics (a few weeks),
where $|\mathcal{D}|$ denotes the number of included outcomes. 
Two kinds of datasets are envisaged, namely,
\textit{training} ($\mathcal{D}_\mathrm{tr}$), used to train prediction models by learning their parameters from real data, 
and \textit{test} ($\mathcal{D}_\mathrm{te}$), used to evaluate their prediction accuracy.

\begin{figure} 
    \begin{center}
    \includegraphics[width=1.0\columnwidth]{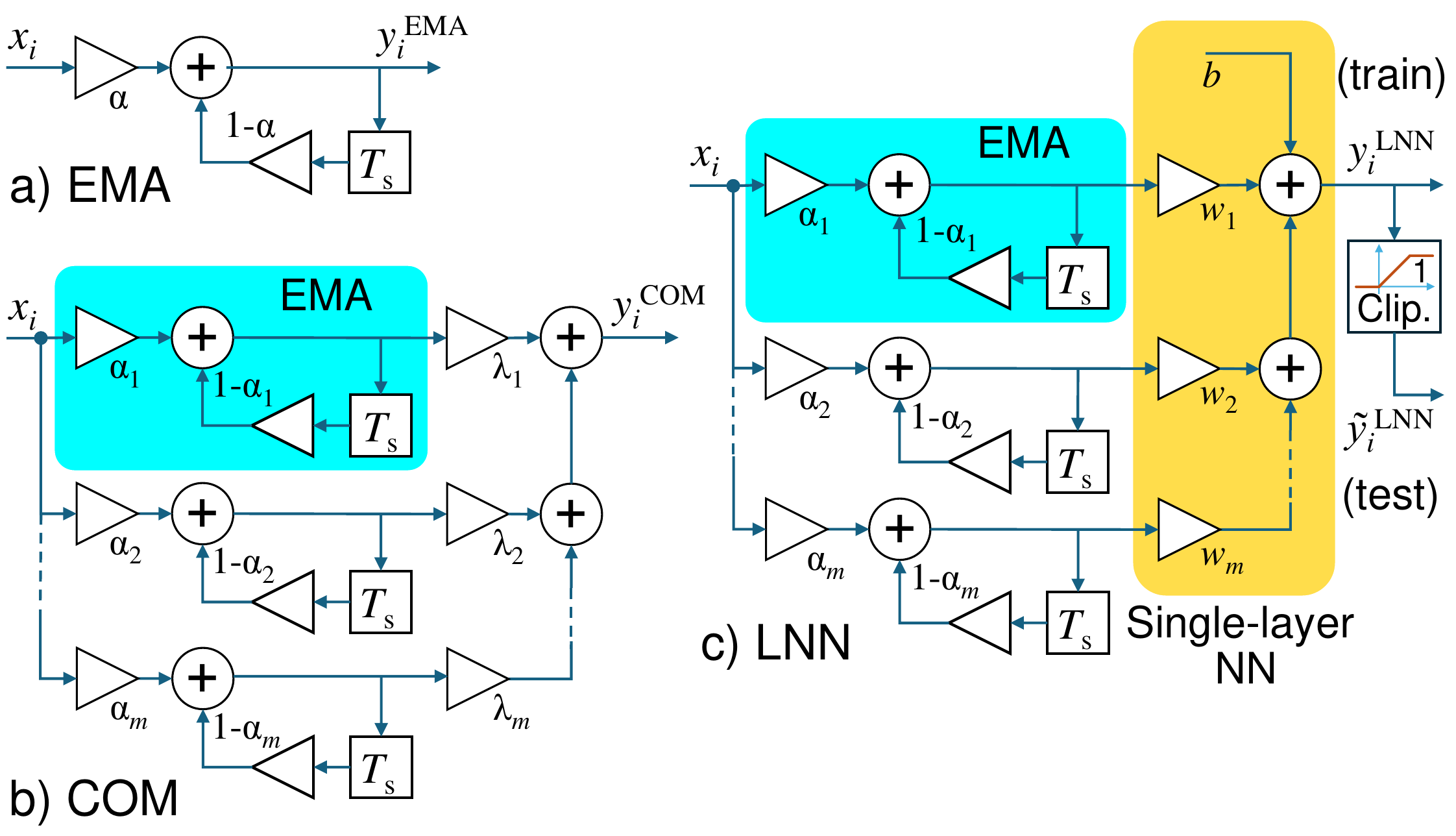}
    \end{center}
    \vspace{-2mm}
    \caption{Operation of prediction models (EMA, COM, and LNN).}
    \vspace{-4mm}
    \label{fig:models}
\end{figure}

The main goal of this work is to find efficient and effective ways to predict the expected quality of the \mbox{Wi-Fi} link over a \textit{reference} interval of a given duration in the near future.
The metric we employ for quality is the FDR, defined as the fraction of transmission attempts the sender performs in said interval that it considers to be successful. 
In the following, a number of simple prediction models are considered, schematically described in Fig.~\ref{fig:models}, and their accuracy is compared.
The simplest one relies directly on the EMA and can be taken as the baseline.
Then, a linear combination of EMA models (COM) was taken into account. 
Finally, an additional method is introduced,
named linear ANN (LNN),
which exploits a simple multilayer perceptron (MLP) with a single layer for combining EMA outcomes 
and enables inexpensive implementations.
Prediction models based on the simple moving average (SMA), as well as on linear and polynomial regression, were not considered since they showed poorer accuracy than EMA \cite{formisPredictingWirelessChannel2023}.
For example, from Table II of that paper, the MSE of the predicted FDR  is $0.987\cdot 10^{-3}$ for SMA, whereas it is $0.803\cdot 10^{-3}$ for EMA, 
which means a $18.64\%$ improvement on accuracy.
Every prediction model $\mathrm{M}$ is characterized by a distinct set $\bm{\psi}_\mathrm{M}$ of parameters, hence it is fully described by the tuple ${\langle \mathrm{M}, \bm{\psi}_\mathrm{M} \rangle}$.

\subsection{Experimental setup}
\label{sec:setup}

\begin{figure} [b]
    \begin{center}
    \vspace{-2mm}
    \includegraphics[width=1.0\columnwidth]{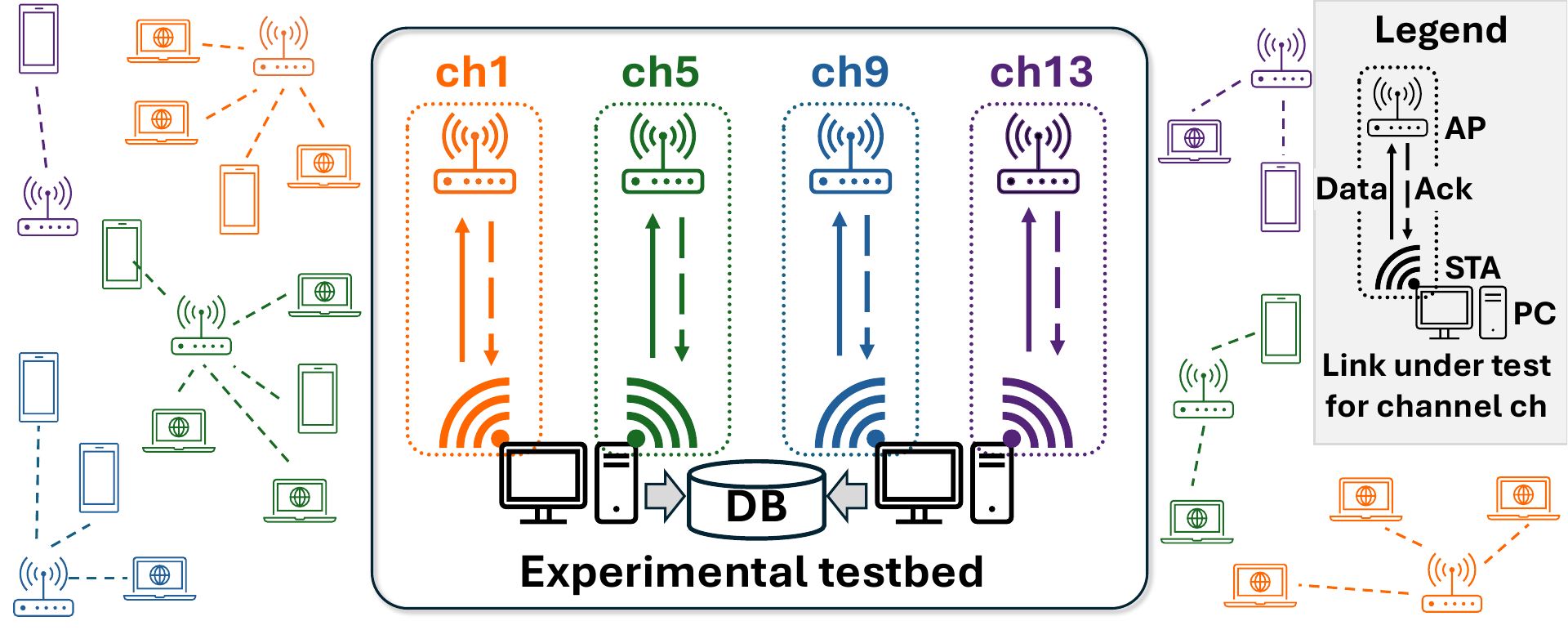}
    \end{center}
    \vspace{-2mm}
    \caption{Experimental testbed: every link under test (operating on channels $1$, $5$, $9$, and $13$) sees a different pattern of interferers.}
    \label{fig:testbed}
\end{figure}

To shorten the time needed to acquire datasets, four non-overlapping channels 
in the \unit[2.4]{GHz} band
were probed contextually by means of two pairs of TP-Link TL-WDN4800 network adapters,
which comply with \mbox{Wi-Fi~4},
installed in two Linux PCs 
(see Fig.~\ref{fig:testbed}).
This is not a limiting choice, as robustness is prioritized over throughput in industrial environments.
Every one of these four STAs was associated to a distinct AP, located about three meters apart from it
(APs were tuned on channels $1$, $5$, $9$, and $13$).
STAs were instructed to repeatedly send in co-ordered way data frames with period $T_\mathrm{s}=\unit[0.5]{s}$ 
and size $\unit[50]{bytes}$ (which is typical of industrial networks).
Having attempts fairly spaced in time is highly desirable, since our prediction models resemble linear digital filters operating on transmission outcomes.
For this purpose, frame aggregation was disabled on the links under test.

Acquisition of datasets relied on a modified version of the \texttt{ath9k} device driver, which permitted to detect the arrival of the ACK frame (or the expiration of the ACK timeout) for 
every transmitted data frame, as well as to disable retransmissions (by setting the retry limit to $0$) and backoff (by setting the contention window to $0$).
By doing so, randomness due to the collision avoidance mechanism of DCF is prevented, and only the effect of disturbance on single transmission attempts was evaluated.
Practically, what we did was to ``sample'' the conditions of every channel in terms of successes and failures of probing attempts on the related link.
Sampling is slow enough ($\unit[2]{Hz}$) so that short error bursts, e.g., due to repeated collisions on air, affect results in a limited way, making outcomes of subsequent attempts reasonably independent:
in fact, every interfering STA terminates any ongoing transmission process
after \texttt{MaxTransmitMsduLifetime} (default value \unit[0.5]{s}).
Besides, by doing so our testbed perturbs the surrounding environment negligibly.
If the sampling rate is increased tangibly, its contribution to channel occupation (which is known) and the related growth of collisions 
(not easy to evaluate)
should be considered as well.

\begin{figure}[b]
    \begin{center}
    \vspace{-3mm}
    \includegraphics[width=\linewidth, keepaspectratio]{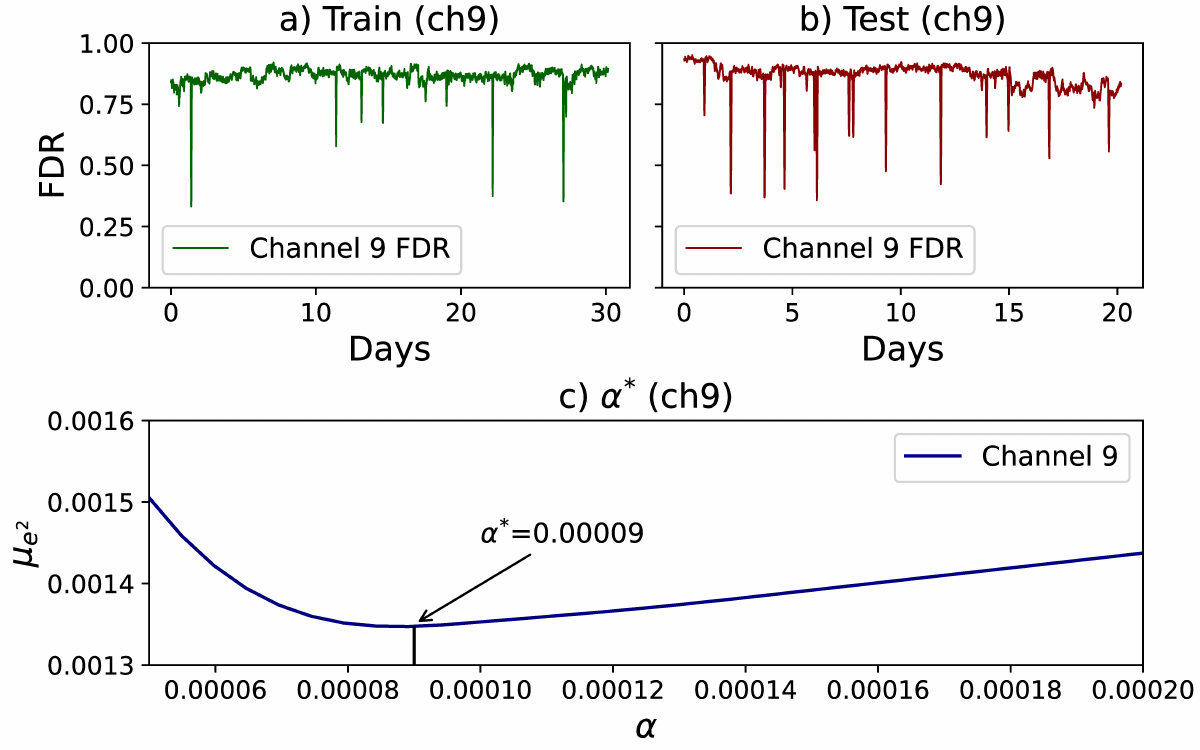}
    \end{center}
    \vspace{-3mm}
    \caption{Timing diagrams for the FDR (target) on channel $9$ used for training (a) and testing (b) models, and MSE vs. pole placement diagram for EMA that highlights the optimal configuration $\alpha^*$ (c).}
    \label{fig:channel_alpha}
\end{figure}

Two experimental campaigns were performed at two distinct times, spaced by a few months, which yielded two sets of four datasets: 
the former, which covered about $30$ days each, were employed for training models; 
while the latter, which covered about $20$ days each, served for testing.
They are denoted $\mathcal{D}_\mathrm{tr}^\mathrm{ch}$ and $\mathcal{D}_\mathrm{te}^\mathrm{ch}$, 
where $\mathrm{ch}\in\{1,5,9,13\}$, and every single dataset includes about $5$ and $3.5$ millions of samples, respectively.
For example, the two diagrams in the upper part of Fig.~\ref{fig:channel_alpha} refer to channel $9$ and report the FDR of the two related datasets,
computed over a $\unit[30]{min}$ moving window centered around the current time $i$,
which can be used as an estimate for the success probability $\varsigma_i$ of the link.
Several disturbance phenomena can be intuitively observed, characterized by different dynamics, which make the FDR keep fluctuating to a certain extent 
(variations are typically less than $\pm~10\%$).
However, from time to time some events occur that cause the link quality to drop abruptly.
The two plots visually bear some resemblance, which highlights the fact that the interference patterns seen on the same channel in the two distinct timeframes were similar.

datasets acquired on the different links are actually representative of distinct spectrum conditions, as:
a) the spatial distribution of nearby interferers (APs and STAs, 
including Wi-Fi 6 ones) 
observed on the related channel is different, as well as the traffic they send on air (amount and pattern); and, 
b) the spatial orientation (angle) of any direct STA$\rightarrow$AP link with respect to walls and objects differed by at least $20^\circ$, which made effects due to multipath fading differ.
Hence, we expect that claims about prediction accuracy derived from them are of quite broad validity.
An informal proof of this can be found in Fig.~3 of \cite{10295470}, which considers datasets acquired with 
a very similar testbed:
as can be seen there, the patterns about the FDR for the four channels differ noticeably, this reflecting the different amount and type of observed interference.

We purposely did not inject any additional interfering traffic, as we are interested in checking to what extent prediction models are able to provide reliable forecasts about the effects of the traffic generated on air by real applications running on nearby nodes (beacons, downloads, multimedia, etc.).
Moreover, in this paper we did not consider the effects of node mobility.
While extremely relevant, this kind of analysis must also consider roaming, and is left as future work together with Wi-Fi network digital twins \cite{scanzioMultiLinkOperationWireless2024}
(it is worth noting that WCP is a basic component of network digital twins and, combined with other elements and methods, it may concur in creating a federated digital twin).
In the absence of mobility, signal attenuation is a lesser cause of unpredictability, 
hence we set the distance between STA and AP similar for all the links.

\subsection{Prediction accuracy}
Since link quality probing occurs at a periodic pace, time proceeds in discrete steps of equal duration $T_\mathrm{s}$.
At any given time $i$, let $x_i$ denote the most recent outcome 
and $y^{\langle \mathrm{M}, \bm{\psi}_\mathrm{M} \rangle}_i$ the FDR \textit{forecast} provided 
by prediction model $\mathrm{M}$ with parameters $\bm{\psi}_\mathrm{M}$.
Clearly, only outcomes $\{ x_l \}_{l \in [1,i]}$ can be used by $\mathrm{M}$ 
to compute $y^{\langle \mathrm{M}, \bm{\psi}_\mathrm{M} \rangle}_i$.
Whatever the model, no predictions are made available until at least
$N_\mathrm{p}$ outcomes have been processed, so that enough information about the link is available to make forecasts accurate.
In this way, the initial transient is skipped and FDR predictions can settle down.

Prediction accuracy at any time $N_\mathrm{p} \leq i \leq |\mathcal{D}|-N_\mathrm{f}$ can be assessed \textit{a posteriori}  using a \textit{target} $z_i$
that coincides with the FDR in a reference interval in the immediate future whose width is $T_\mathrm{f}={N_\mathrm{f} \cdot T_\mathrm{s}}$, 
computed as the SMA of the following $N_\mathrm{f}$ outcomes $\left( x_{i+1}, \cdots, x_{i+N_\mathrm{f}} \right)$ 
\begin{equation}
    \label{eq:target}
    z_i = \frac{1}{N_\mathrm{f}} \sum_{l=i+1}^{i+N_\mathrm{f}}{x_l},
\end{equation}
which provides an unbiased estimate of the mean success probability $\overline\varsigma_i$ in said interval
(we selected $N_\mathrm{f}=3600$, which implies $T_\mathrm{f}=\unit[30]{min}$).
As an example, apart from a $\unit[15]{min}$ time shift, the target for channel $9$ coincides with the plots in Fig.~\ref{fig:channel_alpha}.a/b.
Shrinking $N_\mathrm{f}$ excessively should be avoided, 
since fluctuations 
of the $(x_i)$ random process
may impair the precision of target $z_i$ and render it unreliable:
in fact, it is evaluated as the mean of binary random quantities and its variance depends linearly on $1/N_\mathrm{f}$.
Previous experiments 
on other datasets
showed that the future interval $T_\mathrm{f}$ can be shrunk down to 
$\unit[3]{min}$ 
without impacting on accuracy appreciably, but decreasing it below one minute is not advisable.
Therefore, our prediction models suit the needs of adaptive distributed applications (where, e.g., the rate of non-time-sensitive data exchanges is lowered when spectrum conditions are expected to worsen) but they can hardly be exploited at the MAC layer.
The target is used both in the training phase (where \mbox{$\mathcal{D}=\mathcal{D}_\mathrm{tr}^\mathrm{ch}$}) 
and in the test phase (by setting $\mathcal{D}=\mathcal{D}_\mathrm{te}^\mathrm{ch}$).
In the latter case, it permits to reliably compare the accuracy of different models in the same operating conditions.

Generally speaking, the closer $y^{\langle \mathrm{M}, \bm{\psi}_\mathrm{M} \rangle}_i$ and $z_i$, 
the better the prediction ability of model $\mathrm{M}$ parameterized according to $\bm{\psi}_\mathrm{M}$.
Besides the instantaneous error $e_i = z_i - y_i$, the absolute error $| e_i |$ and the squared error $e_i^2$ were additionally considered.
Starting from them, statistical indices can be evaluated for any given dataset $\mathcal{D}$.
Relevant quantities include average ($\mu_e$, $\mu_{|e|}$, and $\mu_{e^2}$), 
standard deviation ($\sigma_e$, $\sigma_{|e|}$, and $\sigma_{e^2}$),
minimum ($e_{\mathrm{min}}$),
high percentiles ($e_{p90}$, $e_{p95}$, $e_{p99}$, ${|e|}_{p90}$, ${|e|}_{p95}$, ${|e|}_{p99}$,
$e^2_{p90}$, $e^2_{p95}$, and $e^2_{p99}$), 
and maximum ($e_{\mathrm{max}}$, ${|e|}_{\mathrm{max}}$, and $e^2_{\mathrm{max}}$),
every one computed on $|\mathcal{D}|-N_\mathrm{p}-N_\mathrm{f}+1$ predictions, which constitute suitable metrics for assessing prediction accuracy.
For example, the mean squared error (MSE) $\mu_{e^2}$ is customarily employed as the objective function in model training.

For every model we consider, a suitable training phase is preliminarily carried out to determine its optimal parameters.
In particular, we denote $\bm{\psi^*}_\mathrm{M} (\mathcal{D}_\mathrm{tr}^\mathrm{ch})$ the set of parameter values that minimizes $\mu_{e^2}$
for  dataset $\mathcal{D}_\mathrm{tr}^\mathrm{ch}$
\begin{equation}
    \label{eq:ema_opt}
    \bm{\psi^*}_\mathrm{M}  (\mathcal{D}_\mathrm{tr}^\mathrm{ch}) = \arg \min_{\bm{\psi}_\mathrm{M}} 
    \sum_{i=N_\mathrm{p}}^{|\mathcal{D}_\mathrm{tr}^\mathrm{ch}|-N_\mathrm{f}} 
        {\left( z_i - y^{\langle \mathrm{M}, \bm{\psi}_\mathrm{M} \rangle}_i \right)^2 }.
\end{equation}
To achieve a fair comparison, the same training datasets $\mathcal{D}_\mathrm{tr}^\mathrm{ch}$ were used for all models, and above accuracy metrics were evaluated for every optimized model $\langle \mathrm{M}, \bm{\psi^*}_\mathrm{M}  (\mathcal{D}_\mathrm{tr}^\mathrm{ch}) \rangle$ using the very same test datasets $\mathcal{D}_\mathrm{te}^\mathrm{ch}$.
From now on, superscript $^{\langle \mathrm{M}, \bm{\psi}_\mathrm{M} \rangle}$ will be omitted when it can be clearly identified by the context, in particular when referring to optimized models.

\subsection{Exponentially Weighted Moving Average}
Every sequence $(x_i)$ of outcomes can be seen as an instance of a binary random process, which must be filtered suitably to obtain reliable estimates of the instantaneous link quality.
As shown in \cite{FORMIS2024}, one of the most basic (and popular) ways to perform predictions is computing their EMA, described by
\begin{eqnarray}
\label{eq:EMA}
    y_i^{\langle \mathrm{EMA}, \alpha \rangle} = 
    \alpha \cdot x_{i} + (1-\alpha) \cdot y_{i-1}^{\langle \mathrm{EMA}, \alpha \rangle},
\end{eqnarray}
where $y_{i-1}$ denotes the previous prediction and $\alpha$ is a coefficient, also known as smoothing factor, for balancing present and past ($0<\alpha<1$).
Higher values of $\alpha$ make the model more reactive in tracking sudden changes of the link quality, 
while those close to $0$ make it less susceptible to statistical fluctuations. 
Value $y_0$ was initialized to $0.5$, but more fitting estimates should be used if available.

Eq.~\eqref{eq:EMA} corresponds to the simplest form of an infinite impulse response (IIR) low-pass filter
(see Fig.~\ref{fig:models}.a),
described in the \mbox{z-domain} by the transfer function
\begin{align}
\label{eq:Hz}
    H_{\alpha}(z) &= \frac{Y(z)}{X(z)} = \frac{\alpha}{1 - (1-\alpha) z^{-1}} =
    \frac{1-\beta}{1 - \beta z^{-1}},
\end{align}
where $\beta=1-\alpha$ represents the pole.
As can be seen, the EMA model is completely parameterized by the placement of the pole, which implies that $\bm{\psi}_\mathrm{EMA}=\{ \alpha\}$.

Let $\alpha^*$ be the optimal value of $\alpha$ that minimizes the objective function $\mu_{e^2}$ for $\mathcal{D}_\mathrm{tr}^\mathrm{ch}$ according to \eqref{eq:ema_opt}.
This means that $\bm{\psi^*}_\mathrm{EMA} = \{ \alpha^* \}$.
Determining $\alpha^*$ in real operating conditions is what data-driven training is meant to achieve.
For example, Fig.~\ref{fig:channel_alpha}.c (lower part) depicts $\mu_{e^2}$ vs. $\alpha$ for channel $9$.
Determining the position of the minimum can be done easily for the EMA model, since the above function is typically convex.
In fact, lower values of $\alpha$ decrease jitters (and hence the MSE), making results more stable,
but reducing the cutoff frequency too much makes the EMA unable to promptly track variations of the link quality (the MSE increases again, making predictions worse in those non-stationary scenarios we are interested in).
In the case of channel 9, $\alpha^*=0.00009$.

As well known from the literature, the precision (MSE) with which the instantaneous FDR can be evaluated for binary random processes in stationary conditions is
$\varsigma (1-\varsigma) \cdot {\alpha}/({2 - \alpha})$
for $y_i$ (EMA) and ${\varsigma (1-\varsigma)} / {N_\mathrm{f}}$ for $z_i$ (SMA).
By setting $\varsigma=0.8652$
(the mean value of outcomes $x_i$ in $\mathcal{D}_\mathrm{tr}^\mathrm{9}$)
we obtain that the overall MSE is $3.76 \cdot 10^{-5}$ 
(the sets of samples used by EMA and SMA are disjoint, and their MSEs can be summed), 
much smaller than the minimum $\mu_{e^2}$ observed in Fig.~\ref{fig:channel_alpha}.c for real channel $9$ (about $1.35 \cdot 10^{-3}$).
This means that prediction errors mostly depend on sudden variations of the spectrum conditions,
which cannot be foreseen by moving averages, and not on uncertainties due to an insufficient number of samples.

\subsection{Linear Combination of EMAs}
\begin{figure*}[t]
    \begin{center}
    \includegraphics[width=1.0\textwidth]{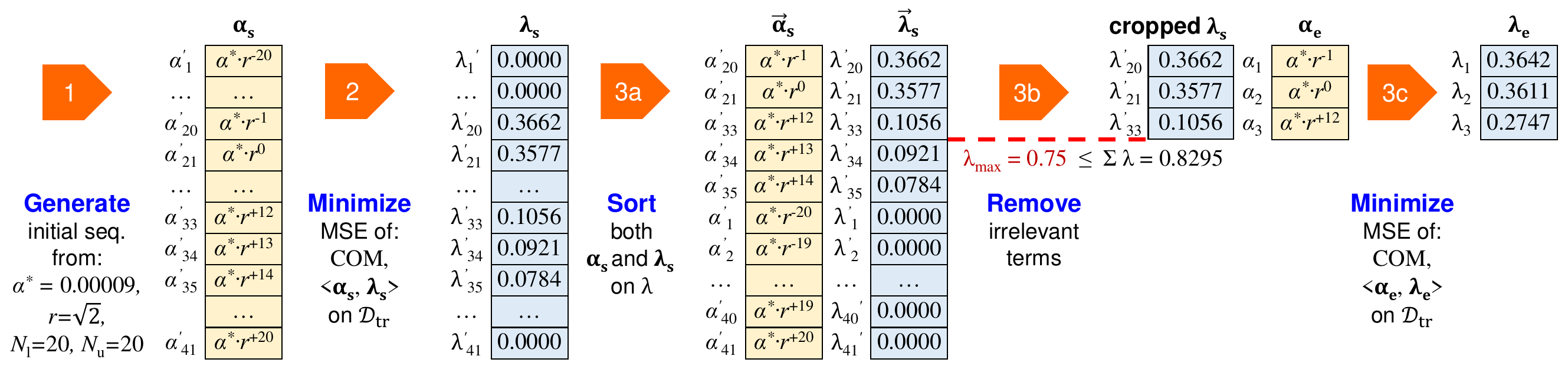}
    \end{center}
    \caption{Training procedure for the COM model (channel $9$ is shown):
    removing irrelevant terms lowers the number of poles from $41$ down to $3$.}
    \label{fig:com}
\end{figure*}

To improve prediction accuracy, a second model (COM) can be exploited, which relies on a linear combination of the output values produced by several concurrently operating EMA models, each one characterized by its smoothing factor $\alpha_j$ (i.e., by the related pole $\beta_j = 1-\alpha_j$),
\begin{eqnarray}
    \label{eq:com_model}
    y^{\mathrm{COM}}_i = \sum_{j=1}^{m} \lambda_{j} \cdot y_i^{ \langle \mathrm{EMA}, \alpha_j \rangle},
\end{eqnarray}
with weights $0 \leq \lambda_{j} \leq 1$ chosen in such a way that
\begin{eqnarray}
    \label{eq:com_constraints}
    \sum_{j=1}^{m} \lambda_{j} = 1.
\end{eqnarray}

The COM model given by \eqref{eq:com_model} is fully described by the tuple 
$ \bm{\psi}_\mathrm{COM} = \langle \bm{\alpha}, \bm{\lambda} \rangle$, where 
$\bm{\alpha}=(\alpha_j)_{j \in [1, m]}$ and $\bm{\lambda}=(\lambda_j)_{j \in [1, m]}$.
It coincides with a multipole low-pass IIR filter (Fig.~\ref{fig:models}.b) characterized by the transfer function
\begin{align}
\label{eq:HzC}
    H_{\langle \bm{\alpha}, \bm{\lambda} \rangle} &= 
    \sum_{j=1}^{m} \lambda_{j} \cdot H_{\alpha_j}(z),
\end{align}
whose poles coincide with vector $\bm{\beta = 1-\alpha}$.
Eqs. (\ref{eq:Hz}) and (\ref{eq:HzC}) in the \textit{z}-domain are not used for calculations (that are carried out directly on samples in the time domain by means of Eqs. (\ref{eq:EMA}) and (\ref{eq:com_model}), for EMA and COM, respectively), 
but are only meant as compact model descriptions.

Unlike the EMA model, finding the optimal configuration $\bm{\psi^*}_\mathrm{COM}$ that minimizes the MSE is not completely trivial.
In the following, a suitable training procedure is described that consists of three steps and provides good results.

\subsubsection{Initial pole selection}
\label{sub:selection}
A number of distinct EMA models are initially considered, 
characterized by different $\alpha'$ coefficients taken from an \textit{initial sequence} defined as
\begin{equation}
    \label{eq:alphaS}
    \bm{\alpha_\mathrm{s}} =
    ( \alpha'_j \mid \alpha'_j=\alpha^* \cdot r^{j-N_\mathrm{l}-1})_{j \in [1, N_\mathrm{l}+N_\mathrm{u}+1]},
\end{equation}
which is a finite-size \textit{geometric progression} with common ratio $r>1$
that includes $|\bm{\alpha_\mathrm{s}}| = N_\mathrm{s} = N_\mathrm{l}+N_\mathrm{u}+1$ elements in strictly increasing order (see block \toeast{1} in Fig.~\ref{fig:com}).
$N_\mathrm{l}$ and $N_\mathrm{u}$ specify the lower and upper bounds for the pool of available $\alpha'$ coefficients, respectively, and are chosen in such a way that the most significant poles of the filter we are looking for fit in the range $[\alpha^* \cdot r^{-N_\mathrm{l}}, \alpha^* \cdot r^{N_\mathrm{u}}]$.
Instead, $r$ affects granularity and permits to adjust the tolerance with which such poles will be approximated.
For example, when $r=\sqrt{2}$, $N_\mathrm{l}=2$, and $N_\mathrm{u}=4$, then
$\bm{\alpha_\mathrm{s}} = ( \frac{\alpha^*}{2}, \frac{\alpha^*}{\sqrt{2}}, \alpha^*, \sqrt{2}\alpha^*, 2\alpha^*, 2 \sqrt{2}\alpha^*, 4\alpha^* )$ and $N_\mathrm{s} = 7$ poles are included.

\subsubsection{MSE minimization}
\label{sub:minimization}
Optimal weights for the above EMA models are then evaluated.
This has been addressed as a minimization problem for the MSE, as formally described by \eqref{eq:ema_opt},
of the COM model specified by \eqref{eq:com_model} over the training dataset $\mathcal{D}_\mathrm{tr}^\mathrm{ch}$, 
constrained by both \eqref{eq:com_constraints} and the given boundaries on every $\lambda_{j}$.
We relied on the Limited-memory Broyden–Fletcher–Goldfarb–Shanno algorithm (\mbox{L-BFGS-B}) \cite{L-BFGS}, 
a quasi-Newton gradient-based optimization method that iteratively updates an estimate of the solution using gradient and curvature information, while also handling boundary constraints. 
It is designed for high-dimensional problems and can find the minimum of a function efficiently. This algorithm is implemented in the \texttt{minimize} function included in the \texttt{Python} library \texttt{scipy.optimize}.
The outcome is a sequence $\bm{\lambda_\mathrm{s}} = ( \lambda'_{1}, \lambda'_{2}, \cdots, \lambda'_{N_\mathrm{s}} )$ of optimal weights for the linear combination in \eqref{eq:com_model},
each one associated to the corresponding smoothing factor in $\bm{\alpha_\mathrm{s}}$ 
(by definition $|\bm{\alpha_\mathrm{s}}| = |\bm{\lambda_\mathrm{s}}| = N_\mathrm{s}$),
which minimizes the MSE for the COM model (block \toeast{2}).
Tuple $\langle \bm{\alpha_\mathrm{s}}, \bm{\lambda_\mathrm{s}} \rangle$
constitutes the initial, optimal selection for parameter configuration.

\subsubsection{Final pole selection}
We observed that a fair number of the weights in $\bm{\lambda_\mathrm{s}}$ evaluated in the previous step were typically rather small, hence the contribution of the related EMA filter to prediction is irrelevant.
This is even more true when test is performed on a dataset other than the one used for minimization, e.g., $\mathcal{D}_\mathrm{te}^\mathrm{ch}$. 
If $\bm{\alpha_\mathrm{s}}$ and $\bm{\lambda_\mathrm{s}}$ are shortened by removing such elements, 
the algorithm becomes faster and more suitable for implementation in embedded devices with scarce computational power, with a negligible impact on accuracy.

The \textit{final sequence} $\bm{\alpha_\mathrm{e}}$ is obtained from $\bm{\alpha_\mathrm{s}}$ by selecting the elements $\alpha'_j$ for which the corresponding weights $\lambda'_j$ are larger.
Practically, let ${\bm{\vec\lambda_\mathrm{s}}} = ( \lambda'_{\pi_1}, \lambda'_{\pi_2}, \cdots, \lambda'_{\pi_{N_\mathrm{s}}} )$ be the same as $\bm{\lambda_\mathrm{s}}$ but sorted in decreasing order,
that is, $\iota < \kappa \implies \lambda'_{{\pi_\iota}} \geq \lambda'_{{\pi_\kappa}}$.
Sequence $(\pi_j)_{j \in [1, N_\mathrm{s} ]}$ is a suitable permutation of $( 1, 2, \cdots, N_\mathrm{s} )$ that enforces such ordering.
Similarly, ${\bm{\vec\alpha_\mathrm{s}}} = ( \alpha'_{\pi_1}, \alpha'_{\pi_2}, \cdots, \alpha'_{\pi_{N_\mathrm{s}}} )$
contains the elements in $\bm{\alpha_\mathrm{s}}$ reordered according to $(\pi_j)$ 
(block \toeast{3a}).
Then, $\bm{\alpha_\mathrm{e}}$ is a prefix of ${\bm{\vec\alpha_\mathrm{s}}}$
\begin{equation}
    \label{eq:selection2}
    \bm{\alpha_\mathrm{e}} = 
    \left( \alpha_j \mid \alpha_j = \alpha'_{\pi_j} \in \bm{\alpha_\mathrm{s}} \right)_{j \in [1, N_\mathrm{e} ]}.
\end{equation}

The number $N_\mathrm{e} = |\bm{\alpha_\mathrm{e}}| \leq N_\mathrm{s}$ of EMA filters that are retained for testing is not fixed.
Instead, it is evaluated as the minimum number of them for which the sum of the related weights exceeds a given threshold $\lambda_\mathrm{max}$ (block \toeast{3b})
\begin{equation}
\label{eq:selection1}
    N_\mathrm{e} = 
    \min \left\{ n \in [ 1, N_\mathrm{s} ]
    \bigg| \sum_{j=1}^{n} 
    {\lambda'_{\pi_j}} \ge \lambda_{\mathrm{max}} \right\}.
\end{equation}
For example, $\lambda_\mathrm{max}=0.75$  means that only the elements strictly needed so that their weights contribute cumulatively for at least $75\%$ are retained.
Setting $\lambda_\mathrm{max}=1.0$ implies that no EMA filters are removed, that is, 
$\bm{\alpha_\mathrm{e}} = \bm{\alpha_\mathrm{s}}$ and $\bm{\lambda_\mathrm{e}} = \bm{\lambda_\mathrm{s}}$.

Since \eqref{eq:com_constraints} must still hold, MSE minimization on $\mathcal{D}_\mathrm{tr}^\mathrm{ch}$ is applied a second time to $\bm{\alpha_\mathrm{e}}$ to obtain the final sequence of weights $\bm{\lambda_\mathrm{e}}$ (block \toeast{3c}). 
Tuple $\bm{\psi^*}_\mathrm{COM} = \langle \bm{\alpha_\mathrm{e}}, \bm{\lambda_\mathrm{e}} \rangle$ 
satisfactorily approximates optimality, and constitutes the configuration of parameters we used for testing.

The above optimization procedure was applied to the training datasets collected from the testbed using $r = \sqrt{2}$, $N_\mathrm{l} = 20$, and $N_\mathrm{u} = 20$.
As depicted in the example of Fig.~\ref{fig:com}, which refers to channel $9$, the initial selection   $\bm{\alpha_\mathrm{s}}$ includes $41$ smoothing factors in the range from 
$8.789\cdot 10^{-8}$ to $9.216\cdot 10^{-2}$.
A first minimization phase on $\mathcal{D}_\mathrm{tr}^\mathrm{ch9}$ provides $\bm{\lambda_\mathrm{s}}$, which is subsequently rearranged in descending order.
By setting $\lambda_\mathrm{max}=0.75$, the number of relevant terms in $\bm{\alpha_\mathrm{e}}$ (i.e., poles of the transfer function) shrinks to just three.
Applying minimization a second time yields $\bm{\lambda_\mathrm{e}}$.

\subsection{Linear combination of EMAs with an ANN}

In this section we explore the integration of EMA and simple neural networks made up of a single linear layer.
This model (LNN) consists of a number of concurrent EMA filters, each one described by \eqref{eq:EMA}, which are fed in parallel with the same outcomes $x_i$ (as in the COM case).
At any discrete time $i$, EMA predictions can be collectively described as a vector
$\mathbf{y}^{\mathrm{EMA}}_i = ( y^{\langle \mathrm{EMA}, \alpha_j \rangle}_i )_{j\in[1,N_\mathrm{s}]}$, where $\alpha_j$ is the $j$-th element in $\bm{\alpha_\mathrm{s}}$
that describes the smoothing factor of the related EMA.
This vector summarizes the information derived from the history of outcomes with a focus on recent observations, but considering different dynamics of disturbance.
Its elements are fed as input features to a single-layer neural network (Fig.~\ref{fig:models}.c), 
whose actual operation is a scalar (dot) product with a \textit{weight} vector $\mathbf{w}$, after which a (scalar) \textit{bias} $b$ is added
\begin{align}
    y^{\operatorname{LNN}}_i = \mathbf{y}^{\mathrm{EMA}}_i \cdot \mathbf{w} + b.
\end{align}

This model, for which $\bm{\psi}_\mathrm{LNN}=\langle \bm{\alpha_\mathrm{s}}, \mathbf{w}, b \rangle$, 
does foresee an identity activation function for training, which makes it completely linear like the COM one.
Unlike the COM model, however, the optimal configuration $\bm{\psi}^*_\mathrm{LNN}$ of the LNN is determined using the conventional ANN training methods based on the Adam optimizer, which was chosen for its ability to dynamically adapt to loss gradients during training, facilitating effective and rapid learning. 
The MSE was selected as the loss function, to assess the discrepancy between model predictions and the target.
To promote convergence towards a global minimum, a gradual reduction of the learning rate was implemented during training.
Additionally, the model's performance was continuously monitored using metrics like MSE and the mean absolute error  (MAE) $\mu_{|e|}$, ensuring effective convergence during training.
More in detail, the model was trained with the \texttt{Keras} module of \texttt{TensorFlow} in $15$ epochs, with batchsize equal to $64$, and the learning rate initialized to $0.01$ and halved at each epoch. Weights were randomly initialized with the \texttt{Glorot normal} initializer.

To always provide meaningful results, a clipping function 
$\tilde{y}^{\operatorname{LNN}}_i = \min(1, \max(0, y^{\operatorname{LNN}}_i) )$
was also defined, 
only used in the test phase, which enforces FDR forecasts to stay in $[0,1]$.

\section{Results}
\label{sec:RESULT}
Experiments were carried out to assess the accuracy of the three proposed prediction models (EMA, COM, and LNN).
Two distinct cases were taken into account for any one of them, which differed for training and, as a consequence, the related optimal configuration parameters $\bm{\psi^*}_\mathrm{M} (\mathcal{D}_\mathrm{tr})$.
The EMA model was taken as the reference, because it offers a good compromise between prediction quality and computational complexity \cite{formisPredictingWirelessChannel2023} (sort of a state-of-the-art).

In the first experimental campaign, \textit{channel-dependent} training was employed for models, 
i.e., the training and test datasets referred to exactly the same channel (we say they are coherent).
While \textit{specialized} training should offer, in theory, the best performance, applying it in real-world scenarios could be cumbersome.
In fact, the training dataset must be acquired from the actual hardware deployed in the intended environment. 
Even worse, training should be periodically repeated to adapt the prediction model to changes in the spectrum conditions. 
This implies that suitable procedures need to be incorporated in \mbox{Wi-Fi} equipment that operate continuously and autonomously, which increase implementation complexity.

Consequently, a second experimental campaign was carried out where above models were treated as \textit{channel-independent} ones,
this meaning that the training and test datasets did not refer to the same channel.
This condition permits to assess models' accuracy against previously unseen interference patterns: 
in fact, different channels 
are characterized by different interfering nodes, 
of different kinds and placed in distinct relative positions, 
and generating different traffic patterns belonging to various protocols (including Bluetooth and wireless sensor networks based on IEEE 802.15.4).
Also their behavior over time can be assumed to vary independently.
The aim is to mimic the behavior of \textit{generalized} training, which can be performed once and for all in the design phase.
Using static, pre-trained models has the advantage that they can be integrated by the manufacturer inside \mbox{Wi-Fi} equipment, which leads to dramatically lower implementation complexity.
Whether or not accuracy of channel-independent models resembles channel-dependent ones is what we wish to determine through experimentation
on real data.

Acquisitions of the test and training datasets were spaced by a few months.
In this time lapse the configuration of most of the nearby APs (not under our control) remained the same: this means that their operating channels and relative positions (distance, angle) with respect to our testbed did not vary.
The same applies to the environment topology, including walls, furniture, and fixed obstacles.
Conversely, 
we expect that
the number of STAs associated to any AP, as well as their position and traffic, 
varies consistently over time,
especially for smartphones and notebooks of students in nearby premises.
This explains why the spectrum conditions we observed on any of the four considered channels differed noticeably from one another, and kept changing over time.

We computed the Pearson correlation coefficients between the outcomes of the transmissions performed at the same time on the different channels and discovered that they were mostly uncorrelated.
Only channels $9$ and $13$ showed some marginal correlation, which was likely due to nearby APs and STAs tuned on ``canonical'' channel $11$, which overlaps with both of them causing contextual interference.
Hence, we deem that:
a) the spectrum conditions described by datasets obtained on distinct channels are different enough and, when they are used to train models, different optimal parameter configurations are expected; and,
b) such datasets are also uncorrelated and can be exploited to perform generalized training as well.

\subsection{Channel-dependent models}

\begin{table*}[t]
\vspace{-4mm}
  \caption{Accuracy of prediction models (EMA, COM, and LNN) with channel-dependent training on channels $1$, $5$, $9$, and $13$.}
  \vspace{-2mm}
  \label{tab:res_depend}
  \footnotesize
  \begin{center}
    \tabcolsep=0.16cm
    \def\arraystretch{1.02}
    \begin{tabular}{ccc|ccc|cccccc|cccc}
    Training / 
    & Prediction & Model  & $\mu_{e^2}$ & $e^2_{\mathrm{p}_{95}}$ & $e^2_{\mathrm{max}}$ & $\mu_{|e|}$ & $\sigma_{|e|}$ & ${|e|}_{\mathrm{p}_{90}}$ & ${|e|}_{\mathrm{p}_{95}}$ & ${|e|}_{\mathrm{p}_{99}}$ & ${|e|}_{\mathrm{max}}$ & ${e}_{\mathrm{min}}$ & ${e}_{\mathrm{p}_{5}}$ & ${e}_{\mathrm{p}_{95}}$ & ${e}_{\mathrm{max}}$ \\
    Test 
    & model (M) & param. ($\bm{\psi^*}_\mathrm{M}$)  & \multicolumn{3}{c|}{$[\cdot 10^{-3}]$} & \multicolumn{6}{c|}{[\%]} & \multicolumn{4}{c}{[\%]} \\
    \hline
    \multirow{4}{*}{\begin{tabular}{l}ch1 /\\ch1
    \end{tabular}} 
        & EMA  & $\alpha^*=0.000900$      & \Br 2.03 & 10.58 & 111.53 & 2.64 & 3.65 & \Br 6.92    & \Br 10.29    & \Br 17.57 & 33.40 & -31.35 & -6.73 & 7.16 & 33.40\\
        & COM  & $m=N_\mathrm{e}=6$         & 1.87 & 9.42 & 117.92 & 2.65 & 3.41 & 6.82         & \Bg 9.70          & 15.81 & 34.33 & -32.64 & -7.07 & 6.57 & 34.34\\
        & COM'  & $m=N_\mathrm{s}=41$ & 1.87 & 9.45 & 117.15 & 2.65 & 3.41 & 6.82      & 9.72       & 15.80 & 34.23 & -32.57 & -7.11 & 6.56 & 34.22\\
        & LNN  & -- 
            & \Bg 1.86 & 9.62 & 112.30 & 2.67 & 3.38 & \Bg 6.78     & 9.81      & \Bg 15.56 & 33.51 & -32.61 & -7.34 & 6.34 & 33.51\\
    \hline
    \multirow{4}{*}{\begin{tabular}{l}ch5 /\\ch5\end{tabular}} 
        & EMA & $\alpha^*=0.000085$     & \Br 1.15 & 2.43 & 190.43 & 1.89 & 2.81 & \Br 3.90     & \Br 4.93      & \Br 12.22 & 43.64 & -43.64 & -3.61 & 4.16 & 15.64 \\
        & COM & $m=N_\mathrm{e}=3$          & 0.96 & 1.91 & 195.45 & 1.73 & 2.57 & 3.41         & 4.38          & 10.28 & 44.21 & -44.21 & -3.40 & 3.41 & 31.42\\
        & COM' & $m=N_\mathrm{s}=41$ & 0.96 & 1.91 & 195.45 & 1.74 & 2.57 & 3.40       & 4.37        & 10.29 & 44.21 & -44.21 & -3.40 & 3.40 & 31.44\\
        & LNN &  -- 
            & \Bg 0.91 & 1.74 & 189.75 & 1.74 & 2.47 & \Bg 3.29     & \Bg 4.17      & \Bg 9.64 & 43.56 & -43.56 & -3.89 & 2.77 & 25.28 \\
    \hline
    \multirow{4}{*}{\begin{tabular}{l}ch9 /\\ch9\end{tabular}} 
        & EMA & $\alpha^*=0.000090$      & \Br 3.68 & 10.03 & 248.20 & 2.65 & 5.46 & \Br 5.54    & \Br 10.01    & \Br 31.44 & 49.82 & -49.82 & -4.56 & 6.01 & 21.43  \\
        & COM & $m=N_\mathrm{e}=3$          & 3.19 & 5.43 & 258.17 & 2.39 & 5.11 & 4.39         & 7.37          & 29.24 & 50.81 & -50.81 & -3.74 & 4.72 & 36.18 \\
        & COM' & $m=N_\mathrm{s}=41$ & 3.19 & 5.40 & 257.83 & 2.39 & 5.12 & 4.37       & 7.35        & 29.28 & 50.78 & -50.78 & -3.73 & 4.71 & 36.27 \\
        & LNN &  -- 
            & \Bg 3.01 & 3.56 & 257.70 & 2.45 & 4.91 & \Bg 4.02     & \Bg 5.97      & \Bg 29.20 & 50.76 & -50.76 & -4.49 & 3.68 & 28.40  \\
    \hline
    \multirow{4}{*}{\begin{tabular}{l}ch13 /\\ch13\end{tabular}} 
        & EMA & $\alpha^*=0.000500$       & \Br 1.22 & 5.82 & 41.55 & 2.42 & 2.53 & \Br 5.51      & \Br 7.63       & \Br 12.29 & 20.38 & -18.45 & -5.31 & 5.76 & 20.38 \\
        & COM & $m=N_\mathrm{e}=6$          & \Bg 1.09 & 4.86 & 40.08 & 2.37 & 2.30 & 5.19      & 6.97       & 11.10 & 20.02 & -16.27 & -5.01 & 5.39 & 20.02 \\
        & COM' & $m=N_\mathrm{s}=41$ & \Bg 1.09 & 4.84 & 40.63 & 2.38 & 2.29 & 5.18    & 6.96     & 11.07 & 20.16 & -16.32 & -5.01 & 5.39 & 20.16 \\
        & LNN & -- 
            & 1.11 & 4.55 & 35.52 & 2.51 & 2.18 & \Bg 5.13          & \Bg 6.74           & \Bg 10.55 & 18.85 & -17.02 & -5.17 & 5.07 & 18.85 \\
    \hline
    \end{tabular}
    \end{center}
\end{table*}
\begin{table*}[t]
    \vspace{-3mm}
    \caption{Accuracy of prediction models (EMA, COM, and LNN) with channel-independent training on channels $1$, $5$, $9$, and $13$.}
  \vspace{-2mm}
  \label{tab:res_independ}
  \footnotesize
  \begin{center}
    \tabcolsep=0.16cm
    \def\arraystretch{1.02}
    \begin{tabular}{ccc|ccc|cccccc|cccc}
    Training / 
    & Prediction & Model  & $\mu_{e^2}$ & $e^2_{\mathrm{p}_{95}}$ & $e^2_{\mathrm{max}}$ & $\mu_{|e|}$ & $\sigma_{|e|}$ & ${|e|}_{\mathrm{p}_{90}}$ & ${|e|}_{\mathrm{p}_{95}}$ & ${|e|}_{\mathrm{p}_{99}}$ & ${|e|}_{\mathrm{max}}$ & ${e}_{\mathrm{min}}$ & ${e}_{\mathrm{p}_{5}}$ & ${e}_{\mathrm{p}_{95}}$ & ${e}_{\mathrm{max}}$ \\
    Test 
    & model (M) & param. ($\bm{\psi^*}_\mathrm{M}$) & \multicolumn{3}{c|}{$[\cdot 10^{-3}]$} & \multicolumn{6}{c|}{[\%]} & \multicolumn{4}{c}{[\%]} \\
    \hline
    \multirow{3}{*}{\begin{tabular}{l}ch1 /\\ $\mathrm{all}$\end{tabular}} 
        & EMA & $\alpha^{*}=0.000325$ & \Br 2.10 & 10.74 & 114.37 & 2.68 & 3.71 & \Bg 6.97 & \Br 10.36 & \Br 18.14 & 33.82 & -32.11 & -6.73 & 7.25 & 33.82\\
        & COM & $m = N_\mathrm{e}=4$ & 1.99 & 10.46 & 111.03 & 2.71 & 3.55 & \Br 7.31 & 10.23 & 16.76 & 33.32 & -32.95 & -7.76 & 6.91 & 33.32\\
        & LNN & -- 
            & \Bg 1.97 & 10.25 & 108.93 & 2.72 & 3.51 & 7.28 & \Bg 10.12 & \Bg 16.48 & 33.00 & -32.99 & -7.88 & 6.77 & 33.00\\
        \cline{2-16}
    \multirow{3}{*}{\begin{tabular}{l}ch1 /\\ $\overline{\mathrm{ch1}}$\end{tabular}} 
        & EMA & $\alpha^{*}=0.000150$ & \Br 2.56 & 13.54 & 115.81 & 2.96 & 4.11 & \Br 8.02 & \Br 11.64 & \Br 20.70 & 34.03 & -33.64 & -8.28 & 7.71 & 34.03\\
        & COM & $m = N_\mathrm{e}=4$ & 2.14 & 11.45 & 111.29 & 2.77 & 3.70 & 7.52 & 10.70 & 17.47 & 33.61 & -33.61 & -8.12 & 7.09 & 33.42\\
        & LNN & -- 
            & \Bg 2.13 & 11.39 & 111.49 & 2.80 & 3.67 & \Bg 7.51 & \Bg 10.67 & \Bg 17.18 & 33.39 & -3339 & -8.32 & 6.97 & 32.86\\
    \hline
    \multirow{3}{*}{\begin{tabular}{l}ch5 /\\ $\mathrm{all}$\end{tabular}} 
        & EMA & $\alpha^{*}=0.000325$ & \Br 1.24 & 2.63 & 200.70 & 1.81 & 3.03 & \Br 3.62 & \Br 5.14 & \Br 14.65 & 44.80 & -44.80 & -3.49 & 3.74 & 36.84\\
        & COM & $N_{\alpha^{*}}=4$ & 1.00 & 1.91 & 193.97 & 1.70 & 2.66 & 3.32 & \Bg 4.38 & 11.11 & 44.04 & -44.04 & -3.30 & 3.33 & 38.37\\
        & LNN & -- 
            & \Bg 0.99 & 1.93 & 192.69 & 1.71 & 2.63 & \Bg 3.31 & 4.39 & \Bg 10.67 & 43.90 & -43.90 & -3.07 & 3.46 & 37.90\\
        \cline{2-16}
    \multirow{3}{*}{\begin{tabular}{l}ch5 /\\ $\overline{\mathrm{ch5}}$\end{tabular}} 
        & EMA & $\alpha^{*}=0.000475$ & \Br 1.31 & 2.62 & 212.68 & 1.80 & 3.14 & \Br 3.56 & \Br 5.11 & \Br 15.26 & 46.12 & -44.95 & -3.50 & 3.61 & 46.12\\
        & COM & $m = N_\mathrm{e}=4$ & 1.06 & 2.22 & 202.16 & 1.80 & 2.71 & 3.54 & 4.72 & \Bg 11.06 & 44.96 & -44.96 & -3.43 & 3.63 & 42.01\\
        & LNN & -- 
            & \Bg 1.02 & 1.99 & 192.73 & 1.73 & 2.69 & \Bg 3.33 & \Bg 4.46 & 11.37 & 43.90 & -43.90 & -2.96 & 3.56 & 41.33\\
    \hline
    \multirow{3}{*}{\begin{tabular}{l}ch9 /\\ $\mathrm{all}$\end{tabular}} 
        & EMA & $\alpha^*=0.000325$ & \Br 4.43 & 18.88 & 251.93 & 2.60 & 6.12 & \Br 4.44 & \Br 13.74 & \Br 34.45 & 50.19 & -50.19 & -3.62 & 5.03 & 45.90 \\
        & COM & $m = N_\mathrm{e}=4$ & 3.43 & 4.96 & 257.83 & 2.39 & 5.35 & 4.03 & \Bg 7.04 & 31.99 & 50.78 & -50.78 & -3.48 & 4.40 & 47.17 \\
        & LNN & $\alpha^*=0.000325$ & \Bg 3.39 & 5.92 & 259.84 & 2.37 & 5.32 & \Bg 3.90 & 7.70 & \Bg 31.58 & 50.97 & -50.97 & -3.50 & 4.21 & 45.64 \\
        \cline{2-16}
    \multirow{3}{*}{\begin{tabular}{l}ch9 /\\ $\overline{\mathrm{ch9}}$\end{tabular}} 
        & EMA & $\alpha^{*}=0.00045$ & \Br 4.74 & 18.64 & 297.14 & 2.61 & 6.37 & \Br 3.95 & \Br 13.65 & \Br 37.61 & 54.51 & -50.43 & -3.47 & 4.38 & 54.51\\
        & COM & $m = N_\mathrm{e}=5$ & 3.59 & 5.73 & 256.26 & 2.39 & 5.49 & 3.87 & \Bg 7.57 & 32.58 & 50.62 & -50.62 & -3.44 & 4.22 & 50.11\\
        & LNN & $\alpha^{*}=0.00045$ & \Bg 3.50 & 6.17 & 260.32 & 2.39 & 5.41 & \Bg 3.82 & 7.86 & \Bg 32.26 & 51.02 & -51.02 & -3.48 & 4.08 & 48.52\\         
    \hline
    \multirow{3}{*}{\begin{tabular}{l}ch13 /\\ $\mathrm{all}$\end{tabular}} 
        & EMA & $\alpha^{*}=0.000325$ & \Br 1.22 & 5.82 & 40.55 & 2.42 & 2.52 & \Br 5.55 & \Br 7.63 & \Br 12.16 & 20.14 & -17.44 & -5.27 & 5.85 & 20.14 \\
        & COM & $m = N_\mathrm{e}=4$ & \Bg 1.11 & 4.83 & 42.32 & 2.45 & 2.27 & \Bg 5.30 & 6.95 & 10.70 & 20.57 & -16.16 & -5.07 & 5.58 & 20.57 \\
        & LNN & $\alpha^{*}=0.000325$ & 1.13 & 4.71 & 39.35 & 2.52 & 2.23 & 5.33 & \Bg 6.86 & \Bg 10.61 & 19.84 & -16.66 & -5.49 & 5.08 & 10.84 \\
    \cline{2-16}
    \multirow{3}{*}{\begin{tabular}{l}ch13 /\\ $\overline{\mathrm{ch13}}$\end{tabular}} 
        & EMA & $\alpha^{*}=0.000275$ & 1.23 & 5.76 & 40.72 & 2.43 & 2.52 & 5.60 & \Br 7.59 & \Br 12.11 & 20.18 & -16.90 & -5.31 & 5.92 & 20.18\\
        & COM & $m = N_\mathrm{e}=2$ & \Bg 1.16 & 4.92 & 38.01 & 2.51 & 2.29 & \Bg 5.43 & \Bg 7.01 & 10.75 & 19.50 & -16.32 & -5.27 & 5.63 & 19.50\\
        & LNN & $\alpha^{*}=0.000275$ & \Br 1.31 & 5.02 & 36.54 & 2.83 & 2.26 & \Br 5.75 & 7.08 & \Bg 10.54 & 19.12 & -17.60 & -6.34 & 4.36 & 19.11\\
    \hline
    \end{tabular}
    \end{center}
    \vspace{-5mm}
\end{table*}

Results about prediction accuracy when channel-dependent training is exploited are reported in Table~\ref{tab:res_depend}.
For every channel, the metrics about accuracy related to the EMA, COM, and LNN models are shown in separate rows.
Two rows are included for 
the linear combination of EMAs, denoted COM', where all $N_\mathrm{s}$ poles in the initial selection are included ($\bm{\psi^*}_\mathrm{COM'} = \langle \bm{\alpha_\mathrm{s}}, \bm{\lambda_\mathrm{s}} \rangle$),
and COM, which just considers the $N_\mathrm{e}$ most significant ones in the final selection ($\bm{\psi^*}_\mathrm{COM}=
\langle \bm{\alpha_\mathrm{e}}, \bm{\lambda_\mathrm{e}} \rangle$).
The most important metric we consider for accuracy is the MSE. 
However, high order percentiles for the absolute error (${|e|}_{\mathrm{p}_{90}}$, ${|e|}_{\mathrm{p}_{95}}$, and ${|e|}_{\mathrm{p}_{99}}$) are also relevant, as they specify an upper bound on the error incurred by $90\%$, $95\%$ and $99\%$ of all the performed predictions, respectively.
Best and worst cases in the table have been highlighted in green and red, respectively.

As can be seen, COM and LNN always behave better than EMA, with LNN often offering the best accuracy.
The two variants of COM practically provided the same accuracy, which means that retaining all the poles in the low-pass filter described by $\bm{\alpha_\mathrm{s}}$ is useless.
For this reason, only significant terms described by $\bm{\alpha_\mathrm{e}}$ will be retained in the following.
Conversely, the LNN has proved able to exploit all terms in $\bm{\alpha_\mathrm{s}}$.
This is probably due to the training procedure for neural networks, which outperformed the L-BFGS-B 
optimization method.
In this case, better accuracy could justify higher implementation complexity, which is a non-negligible result.

\subsection{Sensitivity analysis of parameter selection}

All the presented models are parameterized by poles (described by
$\bm{\alpha}$), whose placement depends on $r$, $N_\mathrm{l}$, $N_\mathrm{u}$, and $\lambda_{\mathrm{max}}$. 
Basically, the latter parameter $\lambda_{\mathrm{max}}$ was selected in order not to worsen COM performance noticeably. 
In fact, results about COM in Table~\ref{tab:res_depend} are mostly the same as COM'.

\begin{figure}[t]
    \begin{center}
    \includegraphics[width=0.9\columnwidth]{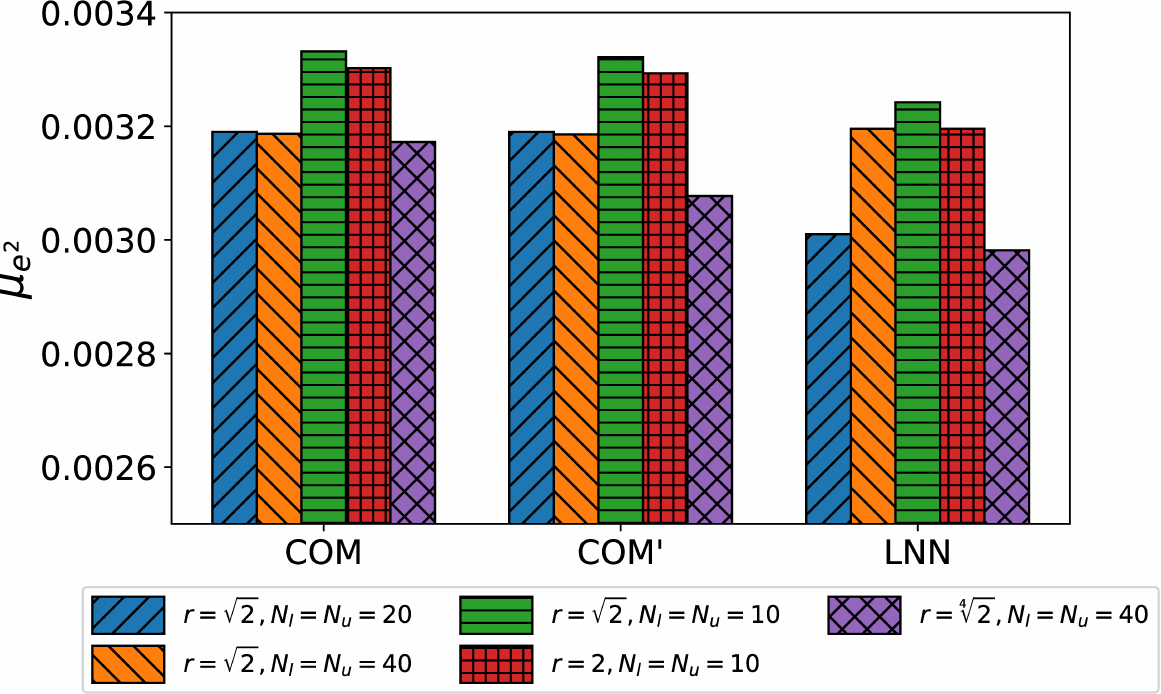}
    \end{center}
    \vspace{-2mm}
    \caption{Parameter sensitivity analysis (MSE vs. smoothing factors).} 
    \label{fig:sensitivity}
    \vspace{-6mm}
\end{figure}

To analyze the sensitivity of training with respect to the model configuration procedure (in terms of the prediction accuracy for the FDR), four additional experiments were carried out 
on channel $9$ 
by varying the former three parameters. 
In the first two experiments we kept the variation step of the smoothing factors $\alpha_j$ fixed ($r=\sqrt{2}$)  and doubled/halved the width of their range by selecting $N_l=N_u=40$ and $N_l=N_u=10$, respectively. 
Results about $\mu_{e^2}$ are sketched in the histogram of Fig.~\ref{fig:sensitivity}, where the first bar of every set of bars (the blue ones) coincides with the value in Table~\ref{tab:res_depend} ($r=\sqrt{2}, N_l=N_u=20$, taken as baseline), while the second and third bars represent two new experimental conditions. 
As can be seen, enlarging the pole range does not lead to better results, while narrowing it is typically pejorative.
In the LNN case, setting $N_l=N_u=40$ leads to a slightly worse accuracy. 
Likely, this is because the newly added poles ($\alpha_j$ values) are not useful for prediction, and so they only have the effect of making convergence of the ANN-based model more difficult.

The latter two experiments are characterized by parameters $r=2, N_l=N_u=10$ and $r=\sqrt[4]{2}, N_l=N_u=40$, respectively. 
They are aimed at investigating the effect of the granularity of smoothing factors $\alpha_j$, leaving their 
range width unchanged with respect to the baseline. 
As expected, lowering the number of poles leads to a degradation of the ability of the model to predict the FDR (last but one bar in every set of bars in Fig.~\ref{fig:sensitivity}). 
Conversely, an increase in their number achieves a slight increase in accuracy (last bars of every set). 
This improvement is, however, irrelevant compared to the additional model complexity, and is inconvenient in many application contexts. 
In conclusion, prediction accuracy loosely depends on the parameters used to determine model configuration,
as long as they are selected near their optimum values.

Regarding LNN, parameters like the learning rate, batch size, and epochs were selected as the result of extensive simulation campaigns aimed at finding the best settings for the training of this kind of models.

\subsection{Channel-independent model and generalization}
Results about prediction accuracy when channel-independent training is exploited are reported in Table~\ref{tab:res_independ}.
For any test dataset (one per channel), two distinct training datasets were considered.
The first, we denote 
``all'', is obtained by merging every training dataset, i.e.,
$
\mathcal{D}_\mathrm{tr}^\mathrm{all} = 
\mathcal{D}_\mathrm{tr}^\mathrm{ch1} \cup
\mathcal{D}_\mathrm{tr}^\mathrm{ch5} \cup
\mathcal{D}_\mathrm{tr}^\mathrm{ch9} \cup
\mathcal{D}_\mathrm{tr}^\mathrm{ch13}$.
This single dataset, which summarizes the behavior of all the considered channels, 
is employed to train all models, reasonably mimicking a \textit{generalized} channel-independent training.

However, some residual correlation exist between $\mathcal{D}_\mathrm{tr}^\mathrm{all}$ and every test dataset $\mathcal{D}_\mathrm{te}^\mathrm{ch}$, as the former includes the training dataset $\mathcal{D}_\mathrm{tr}^\mathrm{ch}$ acquired on the same channel $\mathrm{ch}$.
To remove the effects of such correlation, we performed an additional set of experiments
where the training dataset, we denote $\mathcal{D}_\mathrm{tr}^{\overline{\mathrm{ch}}}$,
explicitly omits the samples of the channel $\mathrm{ch}$ on which test is made, 
in formulas,
$
\mathcal{D}_\mathrm{tr}^{\overline{\mathrm{ch}}} = 
\mathcal{D}_\mathrm{tr}^\mathrm{all} \setminus \mathcal{D}_\mathrm{tr}^\mathrm{ch}
$.
On the one hand $\mathcal{D}_\mathrm{tr}^{\overline{\mathrm{ch}}}$ is completely uncorrelated from
$\mathcal{D}_\mathrm{tr}^\mathrm{ch}$, which makes training \textit{truly generalized}.
On the other, training coverage is worse, as the behavior of one channel was ignored.

Again, the best accuracy is achieved by COM and LNN, the latter showing a slight advantage.
This confirms that, once properly trained, multi-pole IIR low-pass filters manage to improve prediction accuracy tangibly, by tackling the different dynamics of interference on air.

When comparing models by varying the training procedure (specialized, generalized, or truly-generalized), one can see that, as expected, the first provides the best accuracy, as optimization is carried out in conditions that mostly resemble those under which the model is then operated. 
It is also clear that removing the channel under test from training impacts on accuracy negatively.
Therefore, if accuracy is relevant, channel-dependent models are always preferable, although they (repeatedly) require custom training on site.

Interestingly, in the same conditions (same test dataset), both COM and LNN with generalized training behaved slightly better than EMA with a specific training.
This is a very important result, because channel-independent models are way simpler and cheaper than channel-dependent ones, and can be easily included in commercial products.
For example, the generalized COM model we used to test all channels 
in Table~\ref{tab:res_independ} 
has only four poles and is described by attenuations
 $\bm{\alpha} = (8.125\cdot10^{-5}, 5.792\cdot10^{-5}, 1.1483\cdot10^{-4}, 5.201\cdot10^{-3})$
and weights
$\bm{\lambda} = (0.2759, 0.0857, 0.2022, 0.4362)$.
It should be noted that the forecasting error of EMA roughly resembles a SMA evaluated on $N_\mathrm{p}= 2/\alpha-1$ samples.
Hence, the above COM filter approximately behaves as a linear combination of averages computed over four past periods, corresponding to
$24615$, $34532$, $17412$, and $383$ samples, 
with durations of 3h:25m, 4h:48m, 2h:25m, and 3m:11s, respectively.

\begin{figure}[b]
    \begin{center}
    \vspace{-4mm}
    \includegraphics[width=1.0\columnwidth]{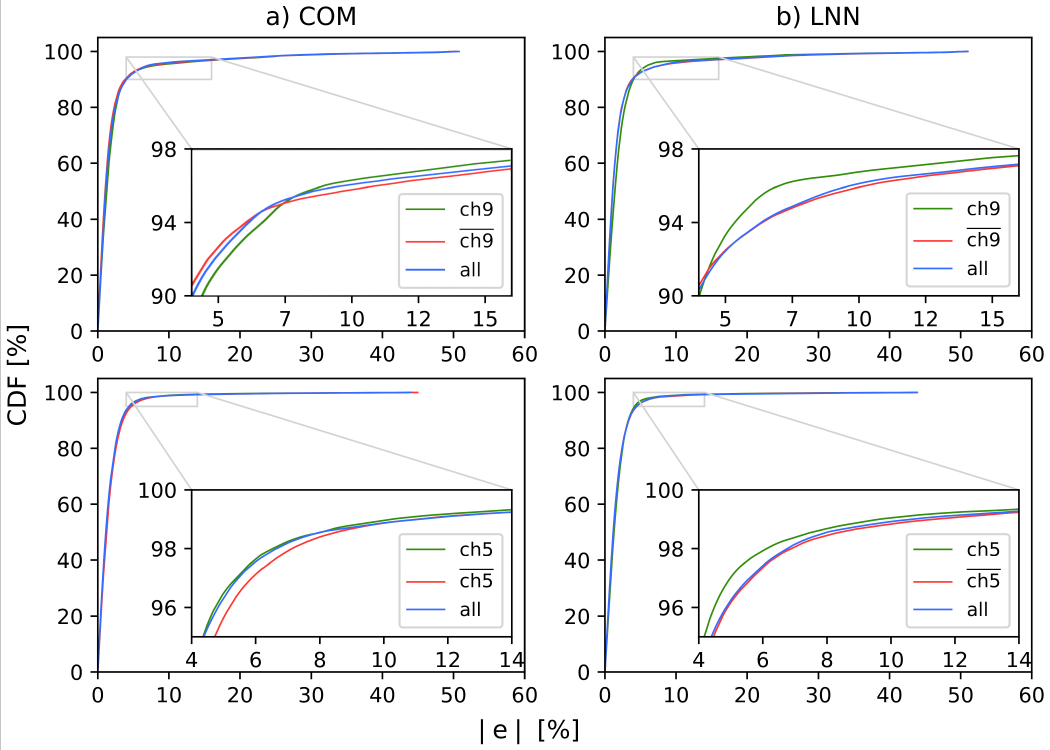}
    \end{center}
    \vspace{-5mm}
    \caption{CDFs of the absolute prediction error $|e|$ on ch. $9$ and $5$ for COM and LNN (channel-dependent/-independent/-truly independent training).}
    \label{fig:plot}
\end{figure}

\subsection{Comparison between COM and LNN}
To better analyze the differences between multi-pole filters (COM and LNN), the cumulative distribution function (CDF) of the absolute prediction error $|e|$ has been computed for the different kinds of training (channel-dependent/-independent/-truly-independent).
We considered channels $9$ and $5$: 
the former was the most problematic one from the point of view of predictability, as witnessed by the high percentiles on the absolute error (${|e|}_{\mathrm{p}_{99}}$), which were by far the worst across all experiments.
By comparison, also channel $1$ often resulted in gross inaccuracies, but not as large as channel $9$.
Channel $5$ showed instead a generally foreseeable behavior concerning most metrics, with the exception of the maximum.

Results are reported in the four
plots of Fig.~\ref{fig:plot}
(channel $9$ in the upper part and channel $5$ below), 
where a specific portion of interest has been zoomed in to provide more details.
The $90$, $95$, and $99$ percentiles correspond to the abscissa of the intersection points between plots and the relevant percentages (in the ordinate).
As can be seen, thanks to the effectiveness of the ANN training technique, which relies on gradient descent and backpropagation, LNN was able to get the most benefits from specialized training performed in the intended operating conditions (on the same channel as test, see the green line),
by lowering the likelihood to make large errors.
COM was seemingly less affected by training, and its accuracy in the channel-dependent case was not remarkably better than in the channel-independent one.
Differences are however small.

It is important to point out that, in harsh conditions like those encountered on channel $9$, 
both COM and LLN provide tangible improvements over EMA:
see, e.g., the $95$ percentile in Tables~\ref{tab:res_depend} and \ref{tab:res_independ}, where the absolute prediction error is always consistently lower, and practically cut in half when solutions based on channel-independent training are exploited.

\subsection{Impact of the future interval}
To provide some insights on the effects of the width $T_\mathrm{f}$ of the reference interval on which the FDR is computed, prediction errors were evaluated on the same datasets 
by choosing $N_\mathrm{f}$ in $\{ 60, 120, 360, 600, 1200, 3600 \}$,
which correspond to durations in the range from $\unit[30]{s}$ to $\unit[30]{min}$.
We considered models with channel-dependent training as they offer the best accuracy.
Results for two channels with quite different behavior ($1$ and $5$), 
which include the optimal attenuation $\alpha^*$ for EMA as well as the MSE $\mu_{e^2}$ 
for both EMA and COM are reported in Table~\ref{tab:future} (LLN is quite similar to COM).

Accuracy of EMA on channel $1$ improves when the reference interval for FDR is shrunk from $\unit[30]{min}$ to $\unit[3]{min}$, since optimization leads to more reactive filters that can track channel quality variations better.
However, it worsen tangibly when the interval reaches $\unit[30]{s}$,
as there are not enough samples and the variability of the random binary process described by outcomes becomes the predominant error source.
Despite the different optima obtained for $\alpha^*$ (especially on large windows), the same behavior is also observed for channel $5$.
Similar results are found for COM, which generally shows better accuracy than EMA but, as expected,
fails to reduce errors when the number of samples is too small ($T_\mathrm{f}=\unit[30]{s}$).

\begin{table}[]
    \caption{Accuracy of EMA and COM vs reference interval width}
    \footnotesize
    \tabcolsep=0.16cm
    \def\arraystretch{1.02}
    \centering
    \begin{tabular}{rr|rrr|rrr}
                   &                & ch1 & \multicolumn{2}{c|}{$\mu_{e^2}$ $[\cdot 10^{-3}]$} & ch5 &                                        \multicolumn{2}{c}{$\mu_{e^2}$ $[\cdot 10^{-3}]$} \\
    $N_\mathrm{f}$ & $T_\mathrm{f}$ & $\alpha^*$ & EMA       & COM    & $\alpha^*$ & EMA       & COM \\
    \hline
          60   & $\unit[30]{s}$    & 0.008120   & \Br 2.404 & \Br 2.400   & 0.007860   & \Br 1.687 & \Br 1.684  \\
         120    & $\unit[1]{min}$   & 0.007480   & 1.627     & 1.622    & 0.007580   & 1.112     & 1.111  \\
         360    & $\unit[3]{min}$   & 0.006280   & \Bg 1.308 & \Bg 1.239    & 0.007040   & \Bg 0.848 & \Bg 0.800  \\
         600    & $\unit[5]{min}$   & 0.005600   & 1.425     & 1.310    & 0.006720   & 0.890     & \Bg 0.806  \\
        1200    & $\unit[10]{min}$ & 0.004640   & 1.863     & 1.602    & 0.006280   & 1.123     & 0.932  \\
        3600   & $\unit[30]{min}$  & 0.000900   & 2.030    & 1.870  & 0.000085   & 1.150    & 0.960 \\
    \end{tabular}
    \label{tab:future}
\end{table}

\subsection{Computational complexity}
The computational complexity of the proposed methods was analyzed in terms of the \textit{training time} for parameterizing the model, the \textit{response time} for its use, and the \textit{memory footprint}.
Experiments reported below were performed on a Linux PC with kernel version 6.8.0-40-generic, equipped with an Intel\textsuperscript{\textregistered} Core\textsuperscript{\texttrademark} i3-10105 CPU running at $\unit[3.70]{GHz}$ and  $\unit[8]{GB}$ of DDR4 RAM. 
Results are reported in Table~\ref{tab:complexity}. 
The training time (not including the evaluation of $\alpha^*$, which is the same for all models) is zero for EMA. 
Training of COM' is faster than COM, as the L-BFGS-B optimizer is invoked just once (and not twice).
By contrast, LNN is the slowest one, with a training time of $\unit[601]{s}$.
To achieve short response times, low computational complexity is a prerequisite for the adoption of the proposed methods in small embedded systems. 
An implementation based on the \texttt{C} programming language and compiled with GCC 11.4.0 without any specific optimization shows that the mean time taken by COM to perform a single prediction (\unit[28]{ns}) is extremely short, especially if compared to what is needed by COM' and LNN ($\unit[179]{ns}$).

Regarding memory footprint, the EMA model only requires two floating points ($\unit[8]{B}$) to store its two parameters (previous prediction $y_{i-1}$ and smoothing factor $\alpha$). 
For the COM and COM' models, memory occupation is $m\cdot\unit[8]{B}+m\cdot\unit[4]{B}$, where $m=N_\mathrm{e}$ for COM and $m=N_\mathrm{s}$ for COM' 
(the second term 
refers to the space 
to store weights $\lambda_j$). 
Finally, the memory footprint for LNN is $m\cdot\unit[8]{B}+(m+1)\cdot\unit[4]{B}$, where $m=N_\mathrm{s}$ (the quantity $m+1$ is the number of weights plus the bias of the output neuron).

\begin{table} 
\caption{Computational complexity of models ($N_\mathrm{e}=6$, $N_\mathrm{s}=41$).}
\centering
\begin{tabular}{l|ccc}
Model & Training time & Mean response time & Memory footprint \\
\hline
EMA & - & \unit[5.6]{ns} & $\unit[8]{B}$ \\
COM & 435 s & \unit[28]{ns} & $\unit[72]{B}$ \\
COM'& 388 s & \unit[179]{ns} & $\unit[492]{B}$ \\
LNN & 601 s & \unit[179]{ns} & $\unit[496]{B}$ \\
\end{tabular}
\vspace{-0.4cm}
\label{tab:complexity}
\end{table}

\subsection{Comparison with deep learning and ARIMA models}
One may wonder how COM and LLN compare to advanced DL models
like convolutional neural networks (CNN), long short-term memory (LSTM),
and bidirectional LSTM (Bi-LSTM).
These models were evaluated in \cite{formis2025improvingwifinetworkperformance} using the same datasets as those we used in this study,
performing both channel-dependent and channel-independent training (not the truly generalized one).
By checking results in that paper against those presented here, it can be seen that prediction accuracy is similar, with EMA-based models often behaving better than DL ones.
This is particularly apparent for channel $13$, characterized by high variability over time, 
and is probably due to the fact that individual (Boolean) transmission outcomes $x_i$ were directly fed as input features in CNN ($3600$ samples) as well as in LSTM and Bi-LSTM ($1200$ samples).
As a consequence, performing training once and for all may result in suboptimal performance.
Conversely, both COM and LNN rely on 
multiple EMA filters, which offer greater adaptability to channel variations.

The key distinction between EMA-based and DL models lies in their computational complexity.
While response time in COM and LLN was always less than $\unit[180]{ns}$, 
a single test operation with DL was many orders of magnitude higher (it took up to $\unit[5.2]{ms}$ for CNN and up to $\unit[78]{ms}$ for the more complex Bi-LSTM).
Also memory footprint differs significantly, with CNN requiring $\unit[30]{kB}$ and Bi-LSTM about $\unit[700]{kB}$.
This makes it clear that, unlike COM and LNN, DL modes can be hardly embedded in inexpensive Wi-Fi equipment.

The classical autoregressive integrated moving average (ARIMA) model was finally evaluated on the same datasets, but results have not been reported because they are sensibly worse than those obtained by the proposed methods.

\section{Conclusions}
\label{sec:CONC}
Data-driven techniques for predicting spectrum conditions in the immediate future with adequate confidence
are among the primary keys for improving dependability and determinism of wireless communication technologies in general, and \mbox{Wi-Fi} in particular.
In fact, information about the current trend of the disturbance affecting a given link (as described by the related FDR) can be exploited, at the MAC and application levels, to counteract its negative effects on communication quality.

Several proposals have appeared on this topic in the past years.
In this paper we explicitly focus on solutions that enable very simple implementations, and can be thus incorporated inexpensively in commercial-off-the-shelf equipment.
In particular, we analyzed three low-pass filters that are fed with the outcomes of transmission attempts.
The first, based on EMA, is characterized by an extremely simple single-pole transfer function, whose cut-off frequency is selected by means of a preliminary data-driven training phase.
The latter two, we term COM and LNN, rely instead on multi-pole transfer functions.
They differ in the way training is carried out.
In the COM model a gradient-based general-purpose technique is applied to minimize the MSE of predictions.
Instead, the LNN model relies on the same training technique used for neural networks.
Experimental results highlight that, most of the times, both COM and LNN provided better accuracy than EMA (LLN behaving slightly better than COM), still remaining simple enough to keep implementation cost low.

Another goal of this research work was to determine how much a specialized training, performed in the expected target operating conditions, improves accuracy over a generalized training.
Results confirmed that the latter approach was slightly worse, but remained nonetheless accurate enough.
This shows that simple prediction models, pre-trained by the producer of \mbox{Wi-Fi} equipment, may offer valuable performance improvements, yet keeping implementation costs low at the same time.
Future work will try to improve accuracy without increasing complexity sensibly, e.g., by including additional input features like 
timings about ACK frame reception and ACK timeout expiration. 
In addition, 
the ability of the models to adapt to the variability of the channel conditions 
(e.g., by using incremental learning and feedback-based model refinement), 
or their use 
on datasets acquired with node mobility,
could be analyzed as well.
Finally, we plan to consider also other classic predictors such as Kalman filtering.

\bibliographystyle{IEEEtran}
\bibliography{BIB_TII-24-5051}

\begin{IEEEbiography}[{\includegraphics[width=1in,height=1.25in,clip,keepaspectratio]{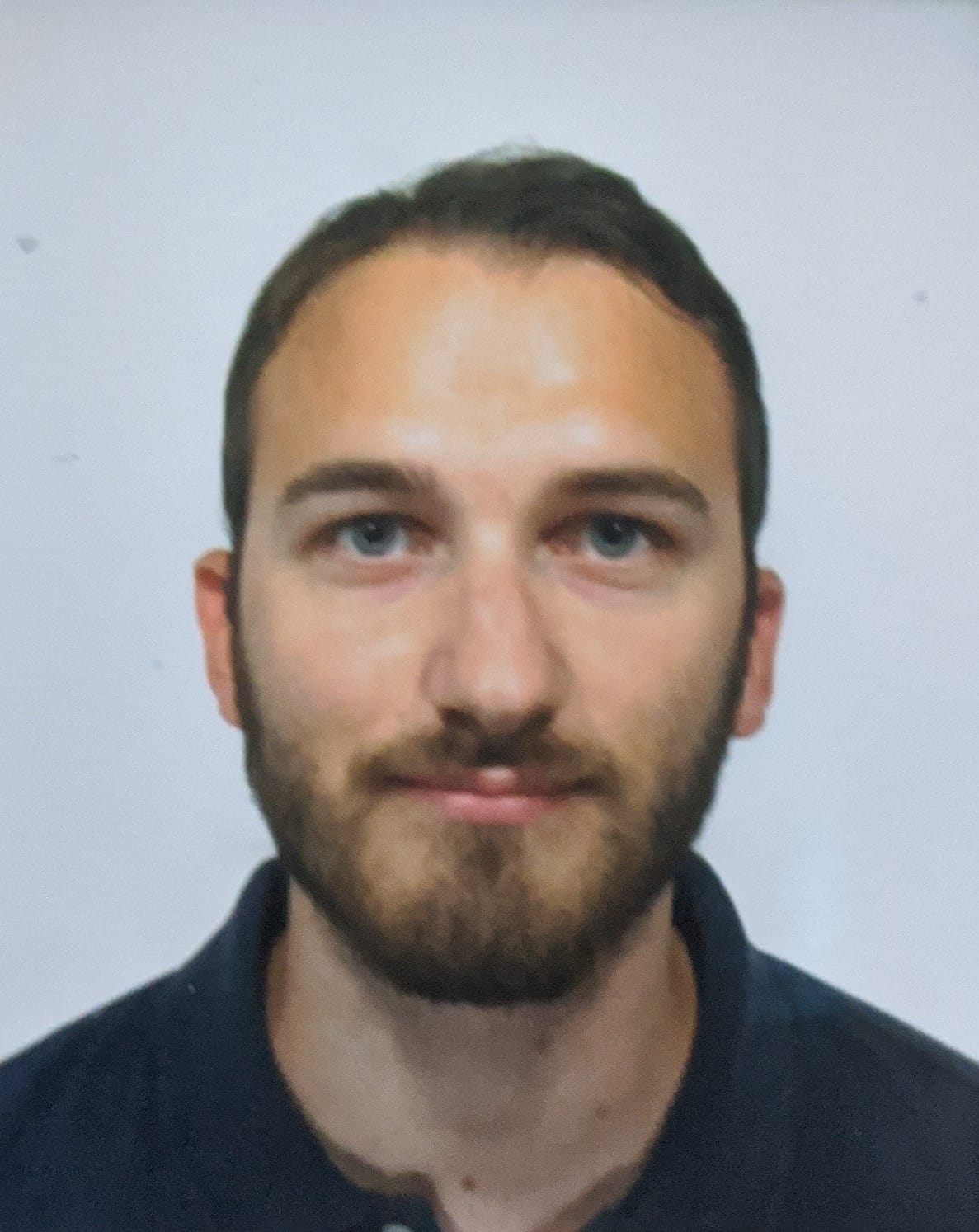}}]{Gabriele Formis} (S'23) received the B.Sc. in mechanical engineering and the M.Sc. degree in automation and control engineering from the Politecnico di Milano, Italy, in 2018 and 2020, respectively. He is currently pursuing the National Ph.D. degree in Artificial Intelligence, Politecnico di Torino, Italy.
In addition, he is Research Associate with the Institute of Electronics, Computer and Telecommunication Engineering of the National Research Council of Italy (CNR-IEIIT). 
His research interests include artificial intelligence, wireless networks, and autonomous driving.
\end{IEEEbiography}

\begin{IEEEbiography}[{\includegraphics[width=1in,height=1.25in,clip,keepaspectratio]{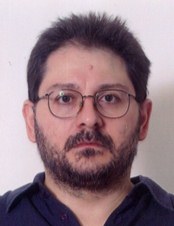}}]{Gianluca Cena}(SM’09) received the M.S. degree in electronic engineering and the Ph.D. degree in information and system engineering from the Politecnico di Torino, Italy, in 1991 and 1996, respectively. 
Since 2005 he has been a Director of Research with the National Research Council of Italy (CNR-IEIIT),
and co-authored about 170 papers and one international patent on
real-time protocols, automotive networks, and wired/wireless industrial communication systems.
He received the Best Paper Award of the IEEE TRANSACTIONS ON INDUSTRIAL INFORMATICS in 2017 and of the IEEE Workshop on Factory Communication Systems in 2004, 2010, 2017, 2019, and 2020. Dr. Cena served as a Program Co-Chairman of the IEEE Workshop on Factory Communication Systems in 2006 and 2008. Since 2009 he has been an Associate Editor of the IEEE TRANSACTIONS ON INDUSTRIAL INFORMATICS.
\end{IEEEbiography}

\begin{IEEEbiography}[{\includegraphics[width=1in,height=1.25in,clip,keepaspectratio]{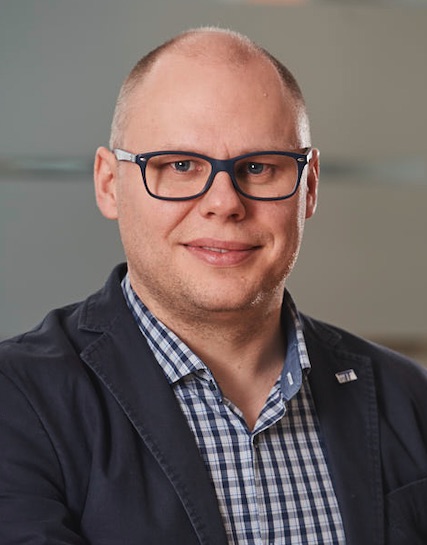}}]{Lukasz Wisniewski} (SM'22) received the Ph.D. (Dr.-Ing.) degree from the Faculty of Integrated Automation, Otto von Guericke University, Magdeburg, Germany.
He completed his informatics studies with the Technical University of Opole, Poland, in 2007. Since 2008, he has been with the Institute Industrial IT (inIT), Technische Hochschule OWL (TH-OWL). Since 2019, he has been an Executive Board Member of inIT and in 2024, he became Deputy Director. In 2022, he was appointed as a Full Professor with TH-OWL, where he chairs the area of technologies of digital transformation. He has coauthored more than 100 articles. He is regularly involved in the organization of several conference bodies of the IEEE Industrial Electronic Society, such as WFCS, ETFA, and INDIN.
\end{IEEEbiography}

\begin{IEEEbiography}[{\includegraphics[width=1in,height=1.25in,clip,keepaspectratio]{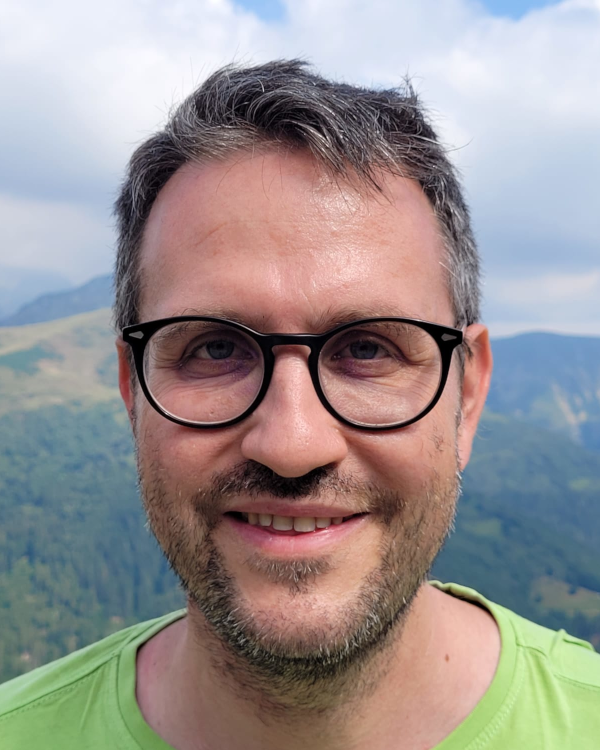}}]{Stefano Scanzio} (S’06-M’12-SM’22) received the Laurea and Ph.D. degrees in computer science from Politecnico di Torino, Turin, Italy, in 2004 and 2008, respectively.
Since 2009, he has been with the National Research Council of Italy, where he is currently a Senior Researcher with the institute CNR-IEIIT. He teaches several courses on computer science. He has authored or coauthored more than 100 papers in international journals and conferences, in the areas of industrial communication systems, real-time networks, wireless networks, and artificial intelligence. He took part in the program and organizing committees of many international conferences of primary importance in his research areas. He received the 2017 Best Paper
Award from IEEE TRANSACTIONS ON INDUSTRIAL INFORMATICS, and four Best Paper Awards in conferences of the IEEE Industrial Electronics Society. He is an Associate Editor of IEEE ACCESS, Ad Hoc Networks (Elsevier), and Electronics (MDPI).
\end{IEEEbiography}

\end{document}